\documentclass[aps,prd,twocolumn,showpacs,amsmath,amssymb,nofootinbib, showpacs, showkeys]{revtex4-2}
\usepackage{graphicx}
\usepackage{color}
\usepackage{times}
\usepackage{inputenc}
\usepackage{bm}
\usepackage{ulem}
\usepackage{multirow}
\usepackage{float}
\usepackage{url}
\usepackage{natbib}
\usepackage{mathrsfs}
\usepackage{physics}
\usepackage{comment}
\usepackage[table,xcdraw]{xcolor}
\usepackage{booktabs,makecell}
\usepackage{comment}
\usepackage[colorlinks=true,citecolor=blue,urlcolor=blue,linkcolor=blue]{hyperref}
\usepackage[caption=false]{subfig}

\usepackage{booktabs}
\usepackage{multirow}

\newcommand{\apjl}{Astro. Phys. J. Lett.}
\newcommand{\mnras}{Mon. Not. Roy. Astron. Soc.}
\newcommand{\aap}{Astronomy $\&$ Astrophysics}
\newcommand{\nphysa}{Nuc. Phys. A}

\begin{document}

\preprint{APS/123-QED}

\title{Exploring $\Delta$-resonance in neutron stars: implications from astrophysical and nuclear observations
}% Force line breaks with \\

\author{Vishal Parmar$^{1}$}
\email{vishal.parmar@pi.infn.it}
\author{Vivek Baruah Thapa$^{2}$}
\email{vivek.thapa@bacollege.ac.in}
\author{Monika Sinha$^{3}$}
\author{Ignazio Bombaci$^{1,4}$}
\affiliation{\it $^{1}$ INFN, Sezione di Pisa, Largo B. Pontecorvo 3, I-56127 Pisa, Italy}
\affiliation{\it $^{2}$ Department of Physics, Bhawanipur Anchalik College, Barpeta, Assam 781352, India}
\affiliation{\it $^{3}$  Indian Institute of Technology Jodhpur, Jodhpur 342037 India}
\affiliation{\it $^{4}$ Dipartimento di Fisica, Universit\`{a} di Pisa, Largo B.  Pontecorvo, 3 I-56127 Pisa, Italy}

\date{\today}% It is always \today, today,

\begin{abstract}
This study presents the first comprehensive Bayesian inference of neutron star matter, incorporating $\Delta$-resonances alongside hyperons and nucleons within a density-dependent relativistic hadron (DDRH) framework. Using constraints from nuclear saturation properties, chiral effective field theory ($\chi$EFT), NICER radius measurements, and tidal deformability data from GW170817, we systematically explore the impact of $\Delta$-resonances on the equation of state (EoS) of dense matter and neutron star observables. Our results demonstrate that the inclusion of $\Delta$-baryons softens the EoS at low densities while maintaining sufficient stiffness at high densities to support $2M_{\odot}$ neutron stars. This naturally reconciles neutron star radius constraints with the recent observation of the low-mass compact object in HESS J1731-347 while simultaneously exhibiting excellent agreement with GW170817 tidal deformability constraints, reinforcing the astrophysical viability of $\Delta$-admixed neutron stars. Additionally, $\Delta$-resonances are found to populate the outer layers of the neutron star core, which may have implications for neutron star mergers and their cooling.  Furthermore, we show that the presence of $\Delta$-baryons might significantly influence neutron star cooling via the direct Urca process. We also investigate quasi-normal $f$-mode oscillations within a fully general relativistic framework, revealing strong correlations between the $f$-mode frequency, neutron star compactness, and tidal deformability. With the inclusion of $\Delta$-resonances and adherence to astrophysical constraints, we obtain $f_{1.4} = 1.97^{+0.17}_{-0.22}$ kHz and the damping time $\tau_{f_{1.4}} = 0.19^{+0.05}_{-0.03}$ s at the $1\sigma$ confidence level.

\end{abstract}

%\keywords{daksjdhankj}%Use showkeys class option if keyword
\maketitle

%\tableofcontents

\section{\label{sec:intro} Introduction}

Neutron stars, among the most extreme objects in the universe, serve as cosmic laboratories that test the limits of physical theories under conditions unattainable on Earth. These stellar remnants have become key to multimessenger astrophysics, with recent breakthroughs from gravitational wave detections of binary neutron star (BNS) mergers \cite{Abbott2017} and precise radius measurements from NICER \cite{Miller2021, Riley_2019} providing unprecedented constraints on the neutron star equation of state (EoS). Additionally, laboratory experiments, such as heavy-ion collisions \cite{PhysRevX.15.021014}, provide complementary insights by probing dense hadronic matter  in terrestrial laboratories. Despite these advances, the internal composition of neutron stars remains an open question.

Their intricate structure spans an outer crust of heavy nuclei, an inner crust with free neutrons and nuclear clusters, and a core where conventional nucleonic matter may give way to more exotic states \cite{Chamel2008, Parmar_2, Glendenning}. The presence of hyperons (e.g. $\Lambda, \Sigma, \Xi$) \cite{Glendenning, Vidana2011, Logoteta2019}, meson condensates (pion and kaon condensates) \cite{KaplanNelson1986, Migdal1978, Glendenning1998}, and resonant baryonic states such as $\Delta$-resonances \cite{Drago2014, Li2019, Sedrakian2022} could dramatically alter the neutron star EoS, influencing its mass-radius relation, tidal deformability, and stability. The transition to quark matter in the core \cite{Alford2005, PhysRevLett.122.061101, Bombaci_2004} or the formation of a color superconducting phase \cite{Rajagopal2001, Lugones_2005} may further modify neutron star properties, potentially leading to hybrid stars with mixed phase interiors. Understanding the role of these exotic components is crucial for refining neutron star models and  the fundamental nature of dense matter.

Although extensive research has explored the presence of hyperons, kaon condensates, and deconfined quarks in neutron stars, relatively few studies have investigated whether $\Delta$ resonances can also exist in these compact stellar objects.  Hyperons emerge at $2$–$3$ times the nuclear saturation density ($\rho_0$), playing a more prominent role in the composition and structure of neutron stars \cite{Bednarek2012, PhysRevC.83.025804, Weissenborn2012}.
The inclusion of hyperons in neutron star matter leads to a significant softening of the EoS, resulting in the well-known ``hyperon puzzle." A similar issue, referred to as the  ``$\Delta$ puzzle" has also been discussed in Ref. \cite{Drago_2014}, raising concerns about how $\Delta$-baryons might further soften the EoS and affect neutron star stability.The seminal works of Glendenning \textit{et al.} \cite{Glendenning1985, Glendenning1998, Glendenning}, 
demonstrated that $\Delta$-resonances are expected to appear only at densities significantly higher than those typically found in the cores of neutron stars, making their presence astrophysically negligible. However, recent studies \cite{Thapa_2021, 2019ApJ...874L..22L} shows $\Delta$ resonance can appear even before two times $\rho_0$, which is well below the typical central density of neutron stars. Even though the appearance of $\Delta$ softens the matter EoS, with density dependent relativistic hadronic model,  massive neutron stars are not ruled out with $\Delta$ particles \cite{2019ApJ...874L..22L}. 

Recently, several studies on $\Delta$-resonances in neutron stars have emerged within the framework of relativistic mean-field (RMF) models \cite{Kolomeitsev2017,Chen2009, Lavagno2010,Schurhoff2010,Drago2016,Drago_2014,Drago2014b,Cai2015,Zhu2016}. Some of these works indicate that $\Delta$-resonances, may appear in nuclear matter at densities as low as $(1-2)\rho_0$ \cite{Cai2015, Drago_2014, Drago2014b, Li2018}, in contrast to earlier studies that predicted significantly higher onset densities. While some investigations neglect hyperons to isolate the effects of $\Delta$-isobars on the nucleonic  EoS \cite{Kolomeitsev2017, Chen2009, Zhu2016}, more recent studies incorporate both hyperons and $\Delta$-baryons, demonstrating that  covariant density functional (CDF) parametrizations satisfying the $2M_{\odot}$ maximum mass constraint remain valid even with the inclusion of $\Delta$-resonances \cite{Li2018, Li2019}. An important consequence of this inclusion is that the resulting neutron star radii shift towards lower values, aligning well with observational constraints on neutron star radii and the tidal deformability constraints from the GW170817 event. These studies suggest the possible existence of ``deltic stars," where up to $20\%$ of the baryons at the stellar core could be $\Delta$-resonances \cite{Marquez2022}. Furthermore, the properties of neutron stars with $\Delta$-admixed hyperonic matter are found to be  highly sensitive to the coupling of $\Delta$-resonances with the $\sigma$ and $\omega$ mesons \cite{Li2018, Cai2015}. Few studies have also explored hypernuclear matter as an intermediate phase between nucleonic and quark matter in the literature \cite{Bonanno2012, Zdunik2013, Dexheimer2015, Drago2014b}. The early onset of $\Delta$ resonances was shown to be essential to ensure the stability of the hadronic star branch \cite{Drago2014b}. All these studies consider a range of $\Delta$-resonance couplings, as their  values remain uncertain due to the lack of direct experimental data. Additionally, a recent study using the quark-meson coupling (QMC) model suggests that the appearance of $\Delta$-isobars may be prohibited \cite{MOTTA2020135266}, further emphasizing the need for a more detailed and systematic understanding of their role in neutron star EoS.

However, because of uncertainties in hyperon interactions, it is possible to fine-tune phenomenological models by calibrating hyperonic interactions to hypernuclear data and ovarious observational constraints from neutron stars, including their masses, radii, tidal deformabilities, and rotational properties. \cite{Bednarek2012, Weissenborn2012, oertel2015hyperons, Tolos2016, Fortin2017}. The addition of $\Delta$-baryons further complicates this problem, as their inclusion might further soften the EoS and potentially make it difficult to support the observational constraint of $\sim 2 M_{\odot}$ neutron stars \cite{lattimer_2011}. Furthermore, the potential interactions of hyperons in dense matter are relatively well-constrained, as they have been extensively studied through hypernuclear experiments and astrophysical observations \cite{Tolos2016}. In contrast, the interactions of $\Delta$-resonances remain significantly more uncertain due to the lack of direct experimental data and their complex coupling to the nuclear medium \cite{Sedrakian2022, Li2018}.

Given the significant uncertainties in $\Delta$-baryon interactions, a systematic study is crucial to understanding their role in dense matter and their impact on neutron star properties. While Bayesian inference has been successfully applied to nucleonic and hyperonic matter to explore the nuclear interaction parameter space and constrain models using astrophysical data \cite{Malik2022, Huang_2024, Sun_2023, Chun_2024}, no such analysis has been performed for $\Delta$-baryons. As a result, their effects on the  EoS, neutron star structure, and compatibility with observational constraints remain largely unexplored in a statistically rigorous framework. Previous studies on the $\Delta$-resonance sector using the relativistic framework have primarily varied the  couplings of $\sigma$-$\Delta$, $\omega$-$\Delta$  and $\rho$-$\Delta$ while keeping the nuclear EoS fixed. This approach restricts a comprehensive understanding of how $\Delta$-baryons influence neutron star properties, as it assumes fixed nuclear matter properties. Consequently, such studies only reveal the effects of $\Delta$-resonances within a predetermined nuclear EoS rather than exploring the full parameter space, which includes nuclear matter properties, hyperons, and $\Delta$-baryons together. A systematic statistical analysis could provide deeper insights into the onset density of $\Delta$-resonances, their impact on EoS , and a more robust determination of $\Delta$-resonance couplings. Additionally, it would clarify their relationship with nuclear matter properties and their compatibility with both neutron star observational data and nuclear matter observables in a unified framework.

This study aims to construct a comprehensive modeling framework using the Density Dependent Relativistic Hadron (DDRH) approach to explore nuclear matter properties, with a specific emphasis on the influence of $\Delta$-resonances in neutron stars. The DDRH  approach has been successfully applied to various nuclear physics problems, including finite nuclei, heavy-ion collisions, and the equation of state of dense matter, demonstrating its effectiveness in describing nuclear interactions across different regimes \cite{DD2, DDME2, typel_2010, Hofmann_2001}.  We incorporate the entire baryon octet alongside $\Delta$-resonances to provide a unified description of dense matter. Our models are constrained using chiral effective field theory ($\chi$EFT), nuclear saturation properties, and astrophysical observations from PSR J0030+0451, PSR J0740+6620 \cite{Miller_2021, Riley_2019}, and the GW170817 event \cite{Abbott2017}. Through Bayesian statistical analysis, we determine $\Delta$-resonance couplings, their impact on the EoS, and their effects on key neutron star properties, including the mass-radius profile, tidal deformability, speed of sound, and their role in the Direct Urca process. Additionally, we assess whether the inclusion of $\Delta$-resonances is consistent with observational constraints from NICER and LIGO/Virgo. We further conduct a correlation analysis to examine the relationships between nuclear matter properties, $\Delta$-resonance couplings, and neutron star characteristics. Furthermore, we perform a non-radial ($f$ and $p1$ mode) oscillation analysis within the framework of full general relativity (FGR), as it provides crucial insights into neutron star interiors, composition, compactness, and underlying nuclear physics \cite{Andersson_1998, guha2023, wen2019, Lau_2024}.

The paper is organized as follows. In Section \ref{formalism}, we briefly discuss the density-dependent relativistic hadron (DDRH) field theory formalism incorporated in this work. We introduce the priors and constraints for our Bayesian analysis. The results are presented in Section \ref{results}, where we discuss the posterior distribution of various nuclear matter and neutron star properties and associated correlations. We summarize our work in Section \ref{summary}.

\section{\label{formalism} Formalism}
\subsection{CDF Model for dense matter}
In this section, we present the density-dependent model for analyzing the appearance of  $\Delta$-resonances  as additional degrees of freedom beside hyperons and nucleons. The matter composition is considered to consist of the baryon octet ($b \equiv N, \Lambda, \Sigma, \Xi$), $\Delta$-resonances ($d \equiv \Delta^{++}, \Delta^{+}, \Delta^{0}, \Delta^{-}$),  alongside leptons ($l$) such as electrons ($e$) and muons ($\mu$).
The strong interactions among baryons are mediated by the scalar $\sigma$, isoscalar-vector $\omega^\mu$, and isovector-vector $\rho^{\mu \nu}$ meson fields, with couplings that evolve as functions of density.  
The hyperon-hyperon interactions are considered to be additionally mediated by the hidden strangeness meson $\phi^\mu$.
It is worth mentioning that the coupling strengths of the $\sigma^*$ meson to hyperons are not well constrained experimentally and are typically inferred from limited experimental data or model-dependent assumptions.
Consequently, in line with previous studies (Ref.-\cite{2022PTEP.2022i3D03M}), we did not take into consideration the $\sigma^*-$hyperon interaction in the present work.
Throughout this model, we adopt natural units ($\hbar = c = 1$). In general, the total Lagrangian density describing the system is given by:\cite{1999PhRvC..60b5803G,2000NuPhA.674..553P,2001PhRvC..64e5805B,2001PhRvC..63c5802B,2000NuPhA.674..553P,1999PhRvC..60b5803G}
\begin{equation}\label{rmftlagrangian}
\begin{aligned}
\mathcal{L} & = \mathcal{L}_B + \mathcal{L}_d + \mathcal{L}_l + \frac{1}{2}(\partial_{\mu}\sigma\partial^{\mu}\sigma - m_{\sigma}^2 \sigma^2) - \frac{1}{4}\omega_{\mu\nu}\omega^{\mu\nu} \\
&  + \frac{1}{2}m_{\omega}^2\omega_{\mu}\omega^{\mu} - \frac{1}{4}\boldsymbol{\rho}_{\mu\nu} \cdot \boldsymbol{\rho}^{\mu\nu} + \frac{1}{2}m_{\rho}^2\boldsymbol{\rho}_{\mu} \cdot \boldsymbol{\rho}^{\mu} -\frac{1}{4}\phi_{\mu \nu} \phi^{\mu \nu}\\
&+\frac{1}{2}m_{\phi}^2 \phi_\mu \phi^\mu,
\end{aligned}
\end{equation}
where $\mathcal{L}_i$ with $i=B,~d,~l$ represent the lagrangian densities of baryon octet, $\Delta-$ baryons and leptonic matter respectively and are given by, 
\begin{equation}
\begin{aligned}
\label{eq:lag}
\mathcal{L}_B & = \sum_{B} \bar{\psi}_B(i\gamma_{\mu} D^{\mu} - m^{*}_B) \psi_B, \\
\mathcal{L}_d & = \sum_{d} \bar{\psi}_d(i\gamma_{\mu} D^{\mu} - m^{*}_d) \psi_d, \\
\mathcal{L}_l & = \sum_{l} \bar{\psi}_l (i\gamma_{\mu} \partial^{\mu} - m_l)\psi_l.
\end{aligned}
\end{equation}
where the fields $\psi_B$, $\psi_{d}$, and $\psi_l$ correspond to the baryon octet, $\Delta$-baryon, and lepton fields respectively.  The covariant derivative  $D_{\mu (j)} = \partial_\mu + ig_{\omega j} \omega_\mu + ig_{\rho j} \boldsymbol{\tau}_j \cdot \boldsymbol{\rho}_{\mu}$, with `$j$' denoting the baryon octet, $\Delta-$ resonances and $\tau_j$  the isospin projections. The interactions between mesons, nucleons, hyperons, and $\Delta-$ resonances are governed by couplings  $g_{ij}$, where $i$ represents the mesons and $j$ denotes the nucleons, hyperons, and $\Delta-$ resonances.  The Dirac effective baryon masses in Eq. \eqref{eq:lag} are given by,

\begin{equation}
    \begin{aligned}
    m^*_B&=&m_B-g_{\sigma B}\sigma\\
    m^*_d&=&m_d-g_{\sigma d}\sigma.
    \end{aligned}
\end{equation}

The ground-state expectation values of various meson fields are discussed in \cite{Thapa_2021, Vivek_2020}. The scalar density and baryon (vector) density for baryons are given by 
$\rho^s= \langle\bar{\psi} \psi \rangle$ and $\rho= \langle\bar{\psi} \gamma^0 \psi \rangle$ respectively.

In the density-dependent model, the meson-nucleon coupling constants vary as a function of density \cite{Malik_2022, Malik_2022_1, Mikhail_2023} as, 
\begin{equation}
\label{eq:density_dependence}
g_{i N}(\rho)= g_{i N}(\rho_{0}) f_i(x), \quad \text{for } i=\sigma,\omega
\end{equation}
where $x=\rho/\rho_0$,  and the density-dependent function $f_i(x)$ is defined as
\begin{equation}
\label{fx}
f_i(x)= \exp(-x^{a_i-1})
\end{equation}
for $\sigma$ and $\omega$ mesons, while for the $\rho$-meson, it is given by
\begin{equation}
\label{fxrho}
f_i(x)= e^{-a_{\rho}(x-1)}.
\end{equation}

%The chemical potential of the $N$-th species, expressed in terms of the Fermi momentum $p_{F_N}$, is given by:
%\begin{equation} \label{eq:chem_pot}
%\begin{aligned}
 %   & \mu_{N} = \sqrt{p_{F_N}^2 + m_{N}^{*2}} + g_{\omega N}\omega_{0} + g_{\rho N} \boldsymbol{\tau}_{N3} \rho_{03} + \Sigma^{r},
%\end{aligned}
%\end{equation}
%where $\Sigma^{r}$ is the rearrangement term introduced to maintain thermodynamic consistency in the DDRH model. It is defined as:
%\begin{equation}\label{eqn.22}
%\begin{aligned}
%\Sigma^{r} & = \sum_{N} \left[ \frac{\partial g_{\omega N}}{\partial \rho}\omega_{0}\rho_{N} - \frac{\partial g_{\sigma N}}{\partial \rho} \sigma \rho_{N}^s + \frac{\partial g_{\rho N}}{\partial \rho} \rho_{03} \boldsymbol{\tau}_{N3} \rho_{N} \right],
%\end{aligned}
%\end{equation}
%where $\rho= \sum_{N} \rho_N$ is the total baryon number density, $\rho_{N}^s$ is the scalar density, and $\sigma$, $\omega_0$, and $\rho_{03}$ are the ground state expectation values of the mesonic fields \cite{Thapa_2021}. This rearrangement term explicitly contributes to the pressure of the matter.

The details of the field equations, energy density, pressure, and other relevant expressions are readily available in the literature. For reference, one can consult \cite{Thapa_2021, Li2019, DDME2, DD2, Malik2022} and the references therein. The onset of $\Delta$-baryons in neutron star matter can occur as either a first-order (spinodal instability) \cite{RADUTA2021136070} or second-order phase transition, depending on the density-dependent interactions of $\Delta$-resonances with nucleons and mesons. If the transition is of first order, a mixed phase is formed where two distinct states coexist: one composed of purely nucleonic matter and the other containing a significant fraction of $\Delta$-baryons. In such a scenario, the Gibbs conditions, along with global baryon number conservation and charge neutrality, must be satisfied to determine the properties of the mixed phase \cite{1999PhRvC..60b5803G,1998PhRvL..81.4564G,1992PhRvD..46.1274G}. For simplicity, in this work, we adopt the maxwell construction as was performed in \cite{RADUTA2021136070}. 
This approach allows for the construction of the EoS for neutron star matter, incorporating the presence of $\Delta$-baryons. To solve the Tolman-Oppenheimer-Volkoff (TOV) equations, this core EoS must be smoothly matched with the EoS of the neutron star crust. In this study, the outer crust EoS is taken from the well-established work of BPS \cite{BPS}, and the inner crust eos is smoothly matched with the core EoS. For more detail please see \cite{Parmar_2024}.

\subsection{Meson-hyperon and meson-$\Delta$ Coupling parameters}\label{sec:Coupling}
For the meson-hyperon vector coupling parameters, we adopt the SU(6) symmetry and the quark counting rule, leading to the following relations:
\begin{equation}
    g_{\omega \Lambda} = g_{\omega \Sigma} = \frac{1}{2} g_{\omega N}, \quad
    g_{\omega \Xi} = \frac{1}{3} g_{\omega N},
\end{equation}
\begin{equation}
    2g_{\phi \Lambda} = 2g_{\phi \Sigma} = g_{\phi \Xi} = -\frac{\sqrt{2}}{3} g_{\omega N},
\end{equation}
\begin{equation}
    g_{\rho \Sigma} = g_{\rho \Xi} = g_{\rho N}, \quad
    g_{\rho \Lambda} = 0.
\end{equation}

The scalar meson-hyperon couplings are determined by considering the hyperon optical potentials, with values set as $U_{\Lambda} = -30$ MeV, $U_{\Sigma} = +30$ MeV, and $U_{\Xi} = -14$ MeV \cite{Thapa_2021}. Due to the limited experimental data on $\Delta$-resonances, the meson-$\Delta$ baryon couplings are treated as free parameters. These couplings are defined as $R_{\sigma \Delta} = g_{\sigma \Delta}/g_{\sigma N}$, $R_{\omega \Delta} =g_{\omega \Delta}/ g_{\omega N}$ and $R_{\rho \Delta} =g_{\rho \Delta}/ g_{\omega N}$. In this work, we consider that the $\rho-$meson to be interacting similarly as for both nucleons as well as non-strange $\Delta-$baryons, so we fix $R_{\rho \Delta}=1$.
Because of the fact that the $\Delta-$resonances do not couple with the strange meson $\phi^\mu$, so we consider $g_{\phi d}=0$.

\subsection{Bayesian Analysis}

Bayesian inference is employed to estimate the model parameters $\mathbf{X}$ by updating prior knowledge using observational as well as experimental data ($D$), following Bayes' theorem:

\begin{equation}
    \label{eq:bayes}
     P(\mathbf{X}|D)=\frac{P(D|\mathbf{X})P(\mathbf{X})}{P(D)}.
\end{equation}

Here, $P(\mathbf{X})$ represents the prior distribution of parameters, updated through the likelihood function $P(D|\mathbf{X})$, while $P(D)$ ensures normalization.
The parameter set $\mathbf{X}$ includes the density-dependent couplings of $\Delta$-resonances in the DDRH framework. Bayesian inference allows for a systematic exploration of parameter uncertainties and their correlations by constraining models against astrophysical and nuclear data. 

To efficiently sample the posterior distribution, we use the nested sampling Monte Carlo algorithm \texttt{MLFriends}, developed by Buchner \cite{Buchner_2016, Buchner_2019}, as implemented in the \texttt{UltraNest} package \cite{Buchner2021}. This method is particularly effective for exploring complex posterior distributions, including those that are multi-modal, exhibit non-linear correlations, and involve large parameter uncertainties. Within \texttt{UltraNest}, the \texttt{Slice sampler} \cite{2019MNRAS.483.2044H} is employed to efficiently navigate high-dimensional parameter spaces while ensuring consistent convergence rates. The number of sampling steps is determined through iterative nested sampling runs, terminating once the natural logarithm of the Bayesian evidence, $\log Z$, reaches convergence.

\subsubsection{Parameters and Priors}
The coupling constants of the DDRH models are treated as free parameters in our Bayesian analysis. The prior distribution for these coupling constants is detailed in Table \ref{tab:prior}. The range for the minimum and maximum values of the coupling constants is informed by Ref. \cite{Malik_2022}, where Bayesian inference was conducted on hyperon signatures within neutron stars using a model akin to the one employed in this study.   The coupling constants between $\Delta$-resonances and mesons remain poorly constrained due to the lack of experimental data. In theoretical studies, these couplings are typically treated as free parameters and varied across different ranges. Some studies adopt fixed values, such as $R_{\omega \Delta} = 1.10$ and $R_{\rho \Delta} = 1$ \cite{Li2018, Li2019}, while others impose conditions where $R_{\sigma \Delta} > R_{\omega \Delta} > 1$ and $R_{\rho \Delta} > 1$ \cite{Marquez2022} or  $0.8 \le R_{\sigma \Delta} \le 1.8$, $0.6 \le R_{\omega \Delta} \le 1.6$ and $0.5 \le R_{\rho \Delta} \le 3$ \cite{RADUTA2021136070, Kolomeitsev2017, Ribes_2019}. Additionally, certain models constrain the difference between scalar and vector couplings within a limited range, ensuring a physically motivated hierarchy. Other studies explore variations with $R_{\rho \Delta} = 1$, $0.8 \leq R_{\omega \Delta} \leq 1.6$, and  $ R_{\sigma \Delta}= R_{\omega \Delta} \pm 0.2$ \cite{Li2019, Li2019a}. Given this wide range of possible values, we adopt $R_{\sigma \Delta}$ within [1.0, 1.5] and $R_{\omega \Delta}$ within [0.8, 1.5], ensuring a broad yet physically motivated parameter space for our analysis. We fix the $R_{\rho \Delta}=1$ for this work as explicitly used in literature \cite{Li2018, Li2019a, Thapa_2021}. On the experimental front, phenomenological model analyses of electron and pion scattering off nuclei, as well as photoabsorption \cite{Drago2014} and $\Delta$ production in heavy-ion collisions \cite{Kolomeitsev2017, COZMA2016166}, provide insights into the $\Delta$ potential. However, there remains no clear consensus on its exact value. Given this uncertainty, we focus on constraining the coupling of $\Delta$-resonances with mesons, treating the $\Delta$ potential as a free parameter. This potential is determined at nuclear saturation density following the standard formulation in terms of the nucleon isoscalar potential. Finally, for each, set of parameter in Table \ref{tab:prior}, we calculate the  hyperon coupling at a fixed value of potential as mentioned in Sec. \ref{sec:Coupling}.
%$U_{\Lambda} = -30$ MeV, $U_{\Sigma} = +30$ MeV, and $U_{\Xi} = -14$ MeV.

\begin{table}[]
\caption{The prior ($P$) configuration used for the parameters of the DDRH model in this study. The terms 'min' and 'max' refer to the lower and upper limits of the considered distribution, respectively.}
\label{tab:prior}
\begin{tabular}{@{}llll@{}}
\toprule
\toprule

Parameters     & \multicolumn{1}{c}{Prior} & Minimum & Maximum \\ \midrule
$M_{\sigma N}$ (MeV) & Fixed                   & \multicolumn{1}{c}{550}     & \multicolumn{1}{c}{550}    \\
$M_{\omega N}$  (MeV)& Fixed                   & \multicolumn{1}{c}{783}     & \multicolumn{1}{c}{783}    \\
$M_{\rho N}$ (MeV)& Fixed                   & \multicolumn{1}{c}{783}     & \multicolumn{1}{c}{783}    \\

$g_{\sigma N}$ & Uniform                   & \multicolumn{1}{c}{8.5}     & \multicolumn{1}{c}{12}    \\
$g_{\omega N}$ & Uniform                   & \multicolumn{1}{c}{9.5}     & \multicolumn{1}{c}{14}    \\
$g_{\rho N}$   & Uniform                   & \multicolumn{1}{c}{2.5}     & \multicolumn{1}{c}{8.0}     \\
$a_\sigma$     & Uniform                   & \multicolumn{1}{c}{0.0}     & \multicolumn{1}{c}{0.20}    \\
$a_\omega$     & Uniform                   & \multicolumn{1}{c}{0.0}     &\multicolumn{1}{c} {0.20}    \\
$a_\rho$       & Uniform                   & \multicolumn{1}{c}{0.0}     & \multicolumn{1}{c}{1.0}     \\
$R_{\sigma \Delta}$ & Uniform  &\multicolumn{1}{c} {1.0}  & \multicolumn{1}{c} {1.5}   \\
$R_{\omega \Delta}$ & Uniform  &\multicolumn{1}{c} {0.8}  & \multicolumn{1}{c} {1.6}   \\ \bottomrule
\end{tabular}
\end{table}

\subsubsection{Constrants}
\textbf{Nuclear matter saturation properties:}
The parameters from the DDRH model are directly linked to the properties of nuclear saturation. Given a specific set of isoscalar and isovector parameters, it becomes possible to calculate several crucial nuclear saturation properties.
The EoS of nuclear matter can be decomposed into two
parts as \cite{Bombaci_1991, malik_2020, Parmar_3}
\begin{equation}
    \epsilon(\rho, \alpha)=\epsilon(\rho, 0) + S(\rho)\alpha^2,
\end{equation}
where $\epsilon$ is the energy per nucleon at a given density $\rho$ and
isospin asymmetry $\alpha=\frac{\rho_n -\rho_p}{\rho_n +\rho_p}$. %\ms{{\Large Is not it $1$?, is it correct definition of $\alpha$?}}. $(\rho)$ \ms{{\Large(Is it $S(\rho)$?)}} 
$S(\rho)$ is defined as the density dependent symmetry energy of the system:
\begin{equation}
    S(\rho)=\frac{1}{2}\Big( \frac{\partial ^2 \epsilon(\rho, \alpha)}{\partial \alpha^2}  \Big)_{\alpha=0}.
\end{equation} 
%\ms{{\Large what is $\delta$? Is it actually $\alpha$ ?}}

%$S(\rho)$ at saturation is one of the most crucial nuclear matter properties and is generally denoted as $J$ or $J_{sym,0}$, known as symmetry energy at saturation. As prescribed in \cite{Glendenning}, the EoS can be expressed using different bulk nuclear matter properties at the saturation density. Specifically, for symmetric nuclear matter, these properties include the energy per nucleon $\epsilon_0 = \epsilon(\rho_0, 0) (n = 0)$, where $\rho_0$ is the saturation density, the incompressibility coefficient $K_0$ (n = 2), the skewness $Q_0$ (n = 3), and the kurtosis $Z_0$ (n = 4).

%\begin{equation}
%    X_0^n=3^n\rho_0^n \Big(\frac{\partial^n \epsilon(\rho, 0)}{\partial \rho^n}\Big)_{\rho_0}; \hspace{1cm} n=2,3,4
%\end{equation}

%Similarly, the symmetry energy can be expanded as a Taylor series around the saturation density $\rho_0$ and the slope $L_{sym,0}$ (n = 1), the curvature $K_{sym,0}$ (n = 2), the
%skewness $Q_{sym,0}$ (n = 3), and the kurtosis $Z_{sym,0}$ (n = 4),
%respectively. These properties of nuclear matter saturation are reasonably well-constrained by experimental data, providing a known plausible range for some of these values. The table \ref{tab:cosntraints} shows the constraints used in the present work.

%\begin{equation}
%    X_{sym,0}^n=3^n\rho_0^n \Big(\frac{\partial^n S(\rho)}%{\partial \rho^n}\Big)_{\rho_0}; \hspace{1cm} n=1,2,3,4.
%\end{equation}

%\ms{{\Large $J_{sym,0}$ appeared in table \ref{tab:cosntraints} is not defined}}

The symmetry energy at saturation, denoted as $J$ or $J_{\text{sym},0}$, is a fundamental nuclear matter property \cite{GARG200736, PhysRevC.76.051603, PhysRevLett.94.032701,Essick_104}. The EoS at saturation density $\rho_0$ can be characterized by bulk nuclear matter properties such as the  energy per nucleon $\epsilon_0$, incompressibility coefficient $K_0$, skewness $Q_0$, and kurtosis $Z_0$. Similarly, the symmetry energy is expanded around $\rho_0$ in terms of its slope $L_{\text{sym},0}$, curvature $K_{\text{sym},0}$, skewness $Q_{\text{sym},0}$, and kurtosis $Z_{\text{sym},0}$. These properties are well-constrained by experimental data, defining a plausible range for nuclear matter parameters. The constraints considered in this study are summarized in Table \ref{tab:constraints}. In this work, we consider two sets of nuclear matter constraints that differ in the range of symmetry energy. Set I  includes constraints derived from purely theoretical approaches or extracted through the theoretical interpretation of experimental data. Set II incorporates the recent PREX-II results on the neutron skin thickness of $^{208}$Pb, which suggest a stiffer symmetry energy \cite{Reed_2021}.
The motivation for using a stiffer symmetry energy range is to explore its impact on $\Delta$-admixed hyperonic neutron stars. Previous studies have indicated that a stiff symmetry energy can lead to a tension between the predicted tidal deformability and the observed values from GW170817 \cite{Reed_2021}. Here, we aim to investigate how this discrepancy manifests in the presence of $\Delta$-baryons in extension with the work as in Ref. \cite{2023EPJWC.27910003T}.

\begin{comment}
    
For a value of certain nuclear matter property (NMP) at the saturation density, we use a probability function ($p_{NMP}$) as follows \cite{Huang_2024}: 
\begin{equation}
\label{eq:nmp_pfunc}
\begin{aligned}
\text{center} &= \frac{NMP_{\text{low}} + NMP_{\text{up}}}{2}, \\
\text{width} &= \frac{NMP_{\text{up}} - NMP_{\text{low}}}{2}, \\
p_{NMP} &= -0.5 \times \frac{\left| \text{center} - NMP \right|^{10}}{\text{width}^{10}}.
\end{aligned}
\end{equation}
Here, $low$ and $up$ represents the lower and upper range of the given nuclear matter property. Such a function is  a super-Gaussian function and is less extreme than a hard
cut, but strongly disfavours values outside the nominal range.
Furthermore, it significantly helps in the e convergence speed of the inference \cite{Chun_2024}.
%\ms{{\Large I think, $p$ and $P$ are same. Make them uniform.}}
\end{comment}

To impose constraints on nuclear matter properties at saturation density, we utilize a probability function that assigns higher likelihood to values within an expected range while strongly suppressing those outside it \cite{Huang_2024, Parmar_2024}. This function, defined as a super-Gaussian ($p_{NMP} = -0.5 \times \frac{\left| \text{center} - NMP \right|^{10}}{\text{width}^{10}}$), determines the probability based on the deviation of a given property from its central value relative to its allowed range, ensuring a smooth transition instead of a hard cut-off. By incorporating this approach, we improve the convergence speed of the inference process \cite{Chun_2024}.

%\ms{{\Large What is the meaning of $N^3LO$ in the table \ref{tab:cosntraints}?}}
\begin{table}[]
\caption{The constraints used in the Bayesian inference of the model parameters to generate the DDRH set with $\Delta$-resonances. These include saturation density $\rho_0$,  energy at saturation $\epsilon_0$, incompressibility $K_0$, symmetry energy $J_{\text{sym},0}$, and energy per particle derived from the $N^3LO$ calculation based on chiral nucleon-nucleon (NN) and three-nucleon (3N) interactions for symmetric (SNM) and pure neutron matter (PNM) \cite{Drischler_2016}. The $N^3LO$ band is expanded by 5\%.}
\label{tab:constraints}
\begin{tabular}{@{}cllll@{}}
\toprule
\toprule

\multicolumn{5}{c}{Nuclear Matter Constraints}                               \\ \midrule
\multicolumn{1}{l}{}    & Parameter    & Unit          & Set I & Set II \\ \midrule
\multirow{5}{*}{SNM}    & $\rho_0$     & $fm^{-3}$     & 0.153 $\pm$ 0.005  & 0.153 $\pm$ 0.005  \cite{TYPEL1999331}   \\
                        & $\epsilon_0$ & MeV         & -16.1 $\pm$ 0.2   & -16.1 $\pm$ 0.2  \cite{Dutra_2014}   \\
                        & $K_0$        & MeV         & 230 $\pm$ 40      & 230 $\pm$ 40   \cite{Rutel_2005}  \\
                        & $J$  & MeV         & 30 $-$ 35 \cite{GARG200736, PhysRevC.76.051603, PhysRevLett.94.032701,Essick_104}    & 33 $-$ 43  \cite{Reed_2021} \\
                        & $\epsilon(\rho)$    & MeV & $N^3LO$        & $N^3LO$  \cite{Drischler_2016}   \\
                        &              &               &                &   \\
\multicolumn{1}{l}{PNM} & $\epsilon(\rho)$    & MeV  & $N^3LO$       & $N^3LO$  \cite{Drischler_2016}    \\ \bottomrule
\end{tabular}
\end{table}

\textbf{Symmetric (SNM) and pure neutron (PNM) matter:} In addition to nuclear matter properties, we incorporate constraints from chiral EFT calculations for symmetric (SNM) and pure neutron matter (PNM) by Drischler \textit{et al.} \cite{Drischler_2016}. Using many-body perturbation theory with seven different Hamiltonians, the authors computed the energy per particle for various isospin asymmetries at low densities. We extract energy values manually from their dataset, sampling at ten equidistant points within the density range of 0.02 to 0.2 fm$^{-3}$. To account for theoretical uncertainties, the $\chi$EFT bands are expanded by 5\%, ensuring compatibility with other \textit{ab initio} calculations. Constraints on pressure are not applied due to uncertainties in the density derivative of the energy per nucleon, as discussed in \cite{Carreau_2019}.

\textbf{GW170817:} GW170817 provides key constraints on the EoS through its tidal deformability measurement. The likelihood function for GW170817 is obtained by interpolating the likelihood distribution provided in \cite{Hernandez_2020}, which was derived from fitting the strain data released by the LIGO/Virgo collaboration. This likelihood is incorporated within the Python package \texttt{toast}\footnote{\url{https://git.ligo.org/francisco.hernandez/toast}} and is expressed as  

\begin{equation}
    \mathcal{L}_{GW170817} = F(\Lambda_1, \Lambda_2, \mathcal{M}, q), 
\end{equation}

where the chirp mass, $\mathcal{M}$, is given by  

\begin{equation}
    \mathcal{M} = \frac{(M_1 M_2)^{3/5}}{(M_1 + M_2)^{1/5}},
\end{equation}

with $M_1$ and $M_2$ representing the masses of the binary components, and $q = M_1/M_2$ denoting the mass ratio. The tidal deformabilities, $\Lambda_1(M_1)$ and $\Lambda_2(M_2)$, characterize how each star deforms under the gravitational field of its companion and depend on the respective masses. To determine these deformabilities, along with the mass and radius of the star, one must simultaneously solve the Tolman-Oppenheimer-Volkoff (TOV) equation and the perturbed tidal field equation \cite{2008ApJ...677.1216H}. This is achieved by integrating both equations from the star’s core to its surface, where the pressure vanishes, given a specific equation of state (EoS) and central pressure.

\textbf{PSR J0030+0451 and PSR J0740+6620:} Precise mass and radius measurements of the pulsars PSR J0030+0451 and PSR J0740+6620, obtained by the NICER collaboration, have placed strong constraints on the equation of state (EoS). For PSR J0030+0451, Riley \textit{et al.} \cite{Riley_2019} reported mass and radius estimates at a 68\% confidence level as $M = 1.34^{(+0.15)}_{(-0.16)}$ $M_\odot$ and $R = 12.71^{(+1.14)}_{(-1.19)}$ km, while Miller \textit{et al.} \cite{Miller_2019} provided an alternative measurement of $M = 1.44^{(+0.15)}_{(-0.14)}$ $M_\odot$ and $R = 13.02^{(+1.24)}_{(-1.06)}$ km.  Similarly, for PSR J0740+6620, Miller \textit{et al.} \cite{Miller_2021} determined the mass and radius to be $M = 2.072^{(+0.067)}_{(-0.066)}$ $M_\odot$ and $R = 12.39^{(+1.30)}_{(-0.98)}$ km, while an alternative analysis reported $M = 2.062^{(+0.090)}_{(-0.091)}$ $M_\odot$ and $R = 13.71^{(+2.61)}_{(-1.50)}$ km \cite{Miller_2021}.  

To incorporate the most recent constraints, we include the latest radio timing measurement of PSR J0740+6620, which estimates a mass of $2.08 \pm 0.07$ $M_\odot$ \cite{2021ApJ...915L..12F}. Additionally, we utilize the ST+PST model samples for PSR J0030+0451 \cite{riley_2019_3386449} and the NICER and XMM observational samples for PSR J0740+6620 \cite{riley_2021_4697625}. These datasets are processed using the kernel density estimation (KDE) method to generate posterior distributions, which are subsequently treated as likelihood functions in our analysis following the approach outlined in \cite{Zhu_2023, Huang_2024}.

Incorporating data from NMP, gravitational waves (GW), and NICER mass-radius measurements, we adopt the following total likelihood function form:

\begin{equation}
    \mathcal{L}(D|X) = \mathcal{L}_{NMP} \times \mathcal{L}_{GW170817} \times \mathcal{L}_{NICER}.
\end{equation}

\subsection{Non-Radial Modes: f- and p1-Mode Oscillations}

In neutron stars, non-radial oscillation modes emerge as a result of disturbances in both the stellar matter and the surrounding spacetime. The characteristic frequencies and damping times of these oscillations offer crucial information about the internal composition and fundamental properties of compact objects. By expressing these perturbations in terms of spherical harmonics, they can be categorized into even and odd parity components. 
Among the non-radial modes, the fundamental $f$-mode and the first overtone p1-mode are particularly relevant. They are the non-radial oscillations with pressure as restoring force.
The $f$-mode has no radial nodes and provides insights into the star's compactness, whereas the p1-mode, with one radial node, is highly sensitive to the internal structure and EoS.
In this study, we focus on such non-radial modes which originate from fluid perturbations that interact with gravitational waves. Specifically, we consider the dominant quadrupolar $(l=2)$ even-parity perturbations using full GR description within the framework of the Regge-Wheeler metric \cite{1967ApJ...149..591T}:

\begin{equation}
    \begin{aligned}
    ds^2=& -e^{2\Phi(r)}[1+r^lH_0(r)\mathcal{Y}_{lm}e^{i\omega t}]dt^2- 2i\omega r^{l+1}H_1(r)\\&\mathcal{Y}_{lm}e^{i\omega t}dt dr+ e^{2\Lambda(r)}[1-r^lH_0(r)\mathcal{Y}_{lm}e^{i\omega t}]dr^2+ \\& r^2[1-r^lK(r)\mathcal{Y}_{lm}e^{i\omega t}][d\theta^2+ \sin^2\theta \ d\phi^2].
\end{aligned}
\end{equation}
Here, $H_0$, $H_1$, and $K$ represent the perturbation functions, while $\mathcal{Y}_{lm}$ denote the spherical harmonics. The complex non-radial mode frequency, $\omega$, consists of a real part corresponding to the oscillation’s angular frequency and an imaginary part whose inverse determines the damping time. The fluid perturbations within the star are described by the fluid Lagrangian displacement vector, given by

\begin{equation}
    \begin{aligned}
    \xi^i=& \{r^{l-1}e^{-\Lambda}W(r), -r^{l-2}V(r)\partial_\theta,-r^{l-2}\sin^{-2}\theta V(r)\partial_\phi\} \\& \mathcal{Y}_{lm}(\theta,\phi) e^{i\omega t}, \label{eq:disp_vec_GR}
\end{aligned}
\end{equation}
 where $W$ and $V$ represent the amplitudes of fluid perturbations. 
At the star’s surface, these fluid perturbations vanish, leaving only space-time perturbations, which can be decomposed into incoming and outgoing gravitational waves at infinity. The complex frequency, $\omega$, is determined as the value for which only the outgoing component remains non-trivial \cite{lindblom1983quadrupole, detweiler1985nonradial}. Several techniques exist to compute non-radial oscillation frequencies, including resonance matching \cite{1969ApJ...158....1T, chandrasekhar1991}, the continued fraction method \cite{sotaniDensityDiscontinuityNeutron2001}, and the WKB \cite{kokkotasWModes1992}. 
In this study, we use direct numerical integration \cite{lindblom1983quadrupole, detweiler1985nonradial} to determine oscillation frequencies and damping times.
For more details, see Refs. \cite{Sotani_2011, thorne1967non, Thakur_2024, Thapa_2023, Thakur_2024_a}

\begin{figure*}[htb!]
    \centering
    \includegraphics[scale=0.25]{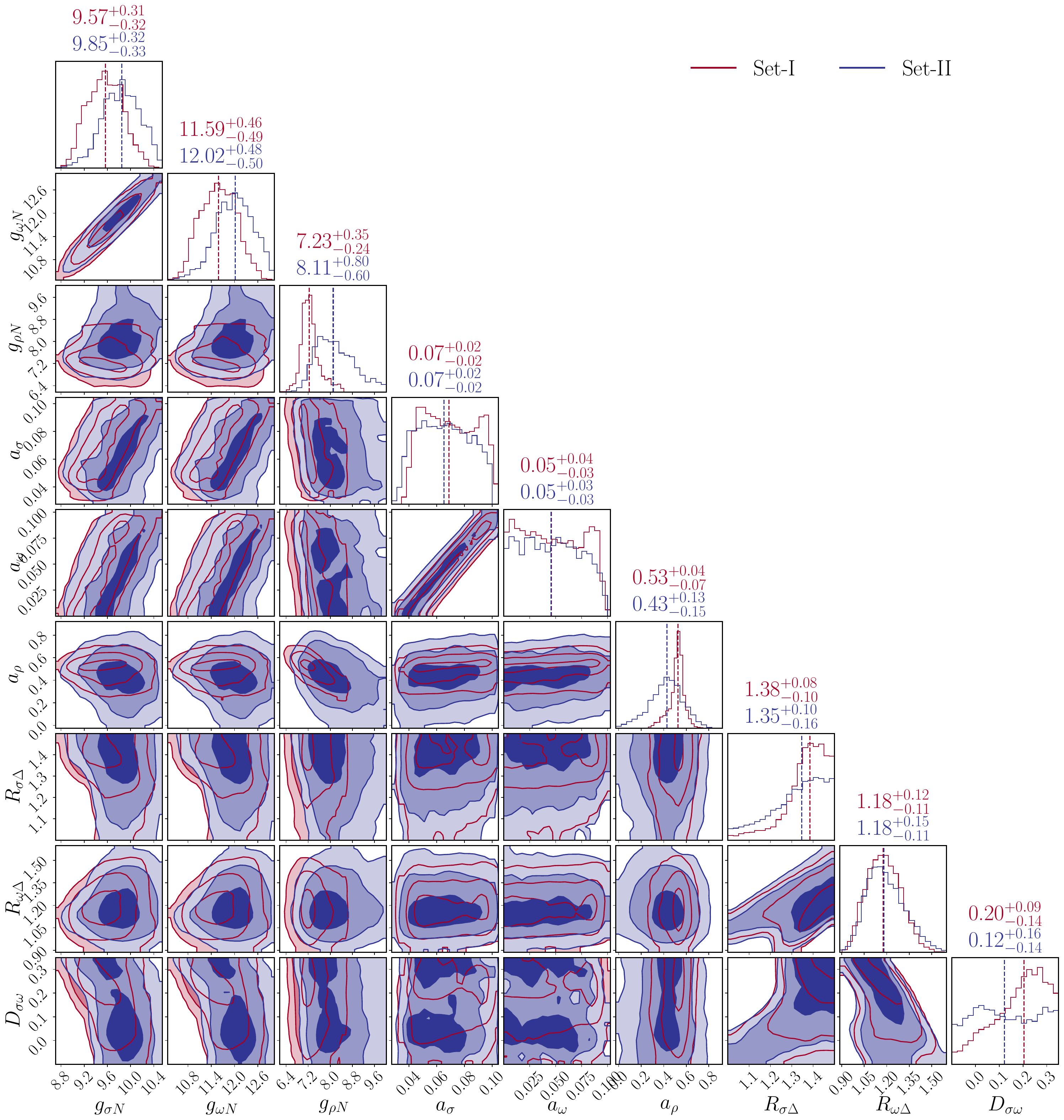}

    \caption{The marginalized posterior distributions of the model parameters. Vertical lines mark the 68\% confidence intervals (CIs). Additionally, the plot includes ellipses representing the 1$\sigma$, 2$\sigma$, and 3$\sigma$ CIs, with darker shades indicating tighter confidence intervals and lighter shades indicating wider intervals in the two-dimensional posterior distributions.}
    \label{fig:par_post}
\end{figure*}

\begin{figure*}
    \centering
    \includegraphics[scale=0.2]{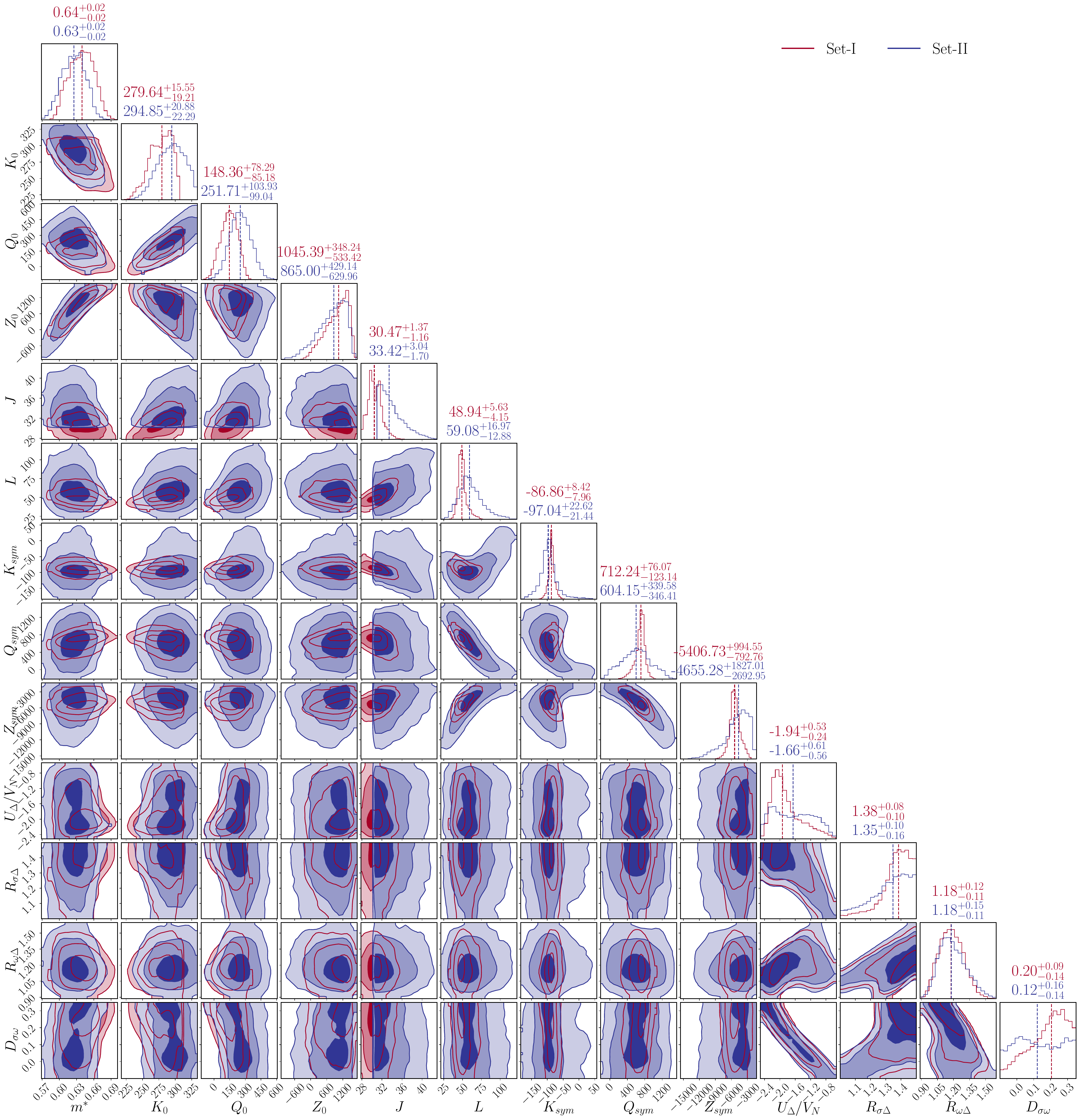}

    \caption{The marginalized posterior distributions of the nuclear matter properties of the parameter sets in Fig. \ref{fig:par_post}. Vertical lines mark the 68\% CIs. Additionally, the plot includes ellipses representing the 1$\sigma$, 2$\sigma$, and 3$\sigma$ CIs, with darker shades indicating tighter confidence intervals and lighter shades indicating wider intervals in the two-dimensional posterior distributions.}
    \label{fig:nuc_prop}
\end{figure*}

\begin{figure*}
    \centering
    \includegraphics[scale=0.22]{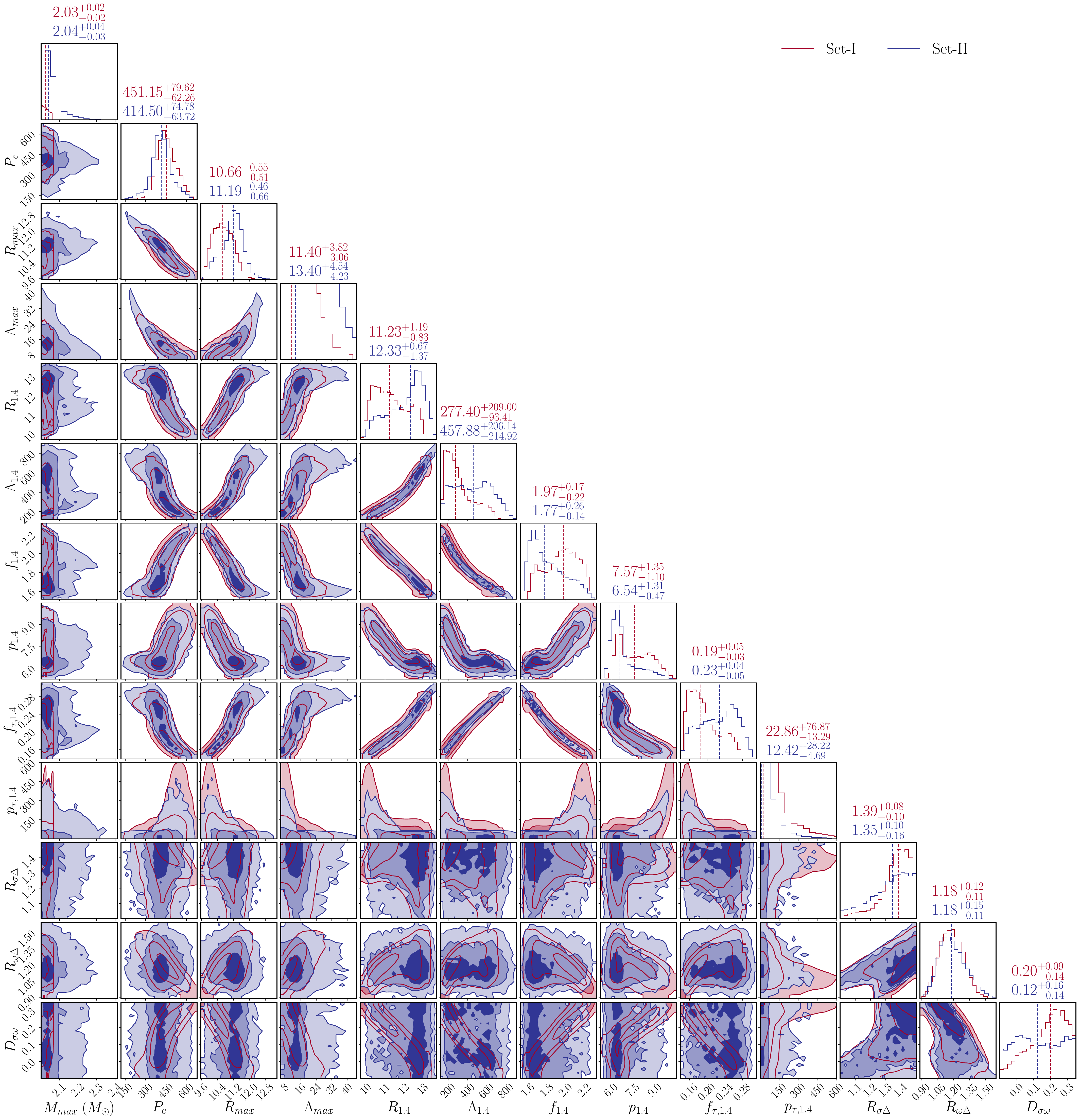}

    \caption{The marginalized posterior distributions of NS properties, including the maximum mass ($M_{\text{max}}$),  central pressure ($P_c$) in MeV-fm$^{-3}$, radius at maximum mass ($R_{\text{max}}$) in km, tidal deformability at maximum mass ($\Lambda_{\text{max}}$), radius of the canonical $1.4 M_\odot$ star ($R_{1.4}$) in km, and its tidal deformability ($\Lambda_{1.4}$), $f_{1.4}$ and $p_{1.4}$ mode frequency in $kHz$ and their respective decay time in Sec.  are shown along with the distribution of  $R_{\sigma \Delta}$, $R_{\omega \Delta}$ and  $D_{\sigma \omega}$. The 2D ellipses and vertical lines have the same interpretation as in Fig. \ref{fig:par_post}.}
    \label{fig:tov_post}
\end{figure*}

\begin{figure}
    \centering
    \includegraphics[scale=0.34]{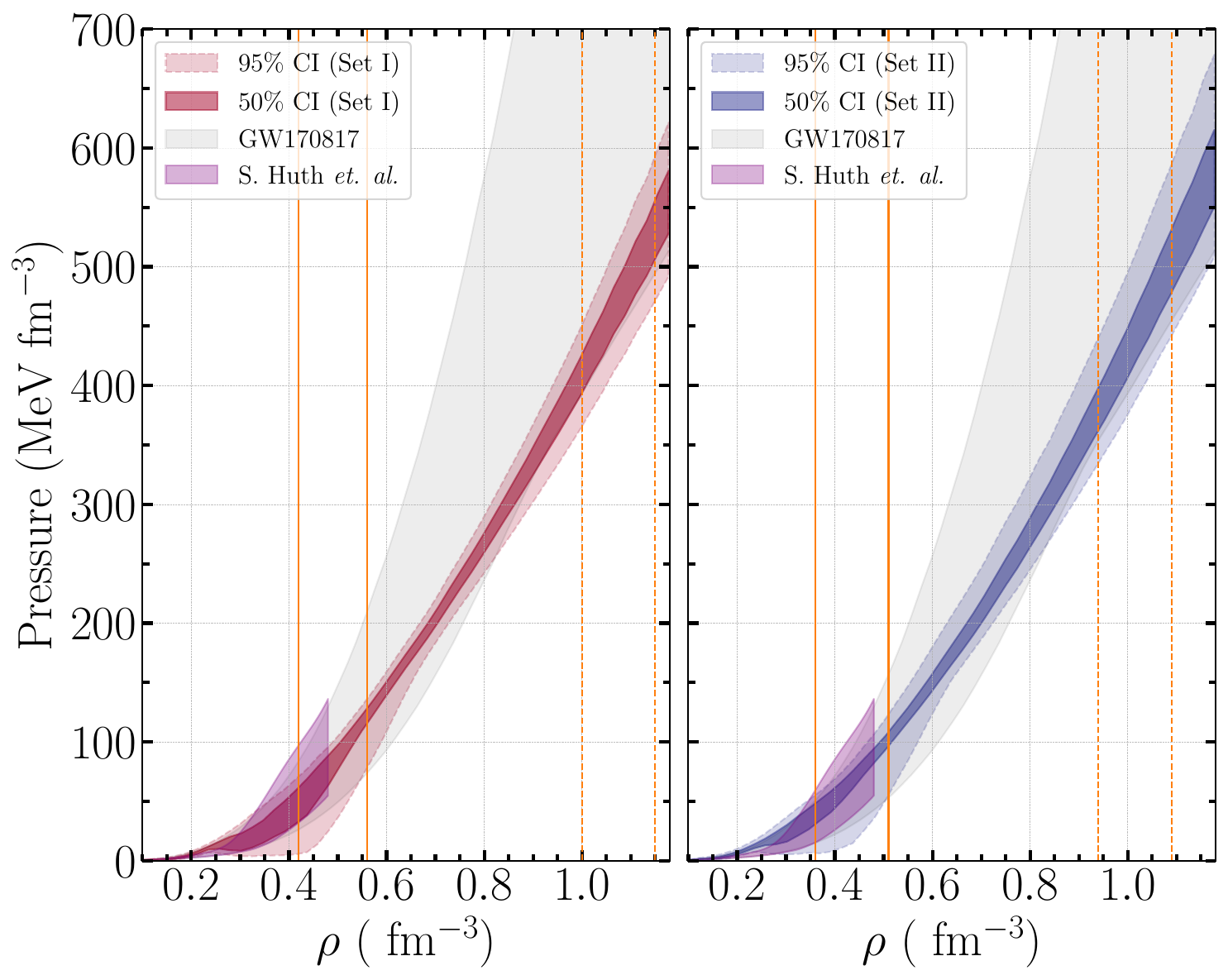}
\caption{The 50\% and 95\% credible intervals (CI) for the EoS of neutron star matter with $\Delta$-resonance admixed hyperonic components. The EoS is compared with the recent estimations from a combination of microscopic nuclear theory and multi-messenger astrophysics \cite{Huth2022}, as well as constraints from GW170817. The vertical solid and dashed lines represent the central densities of a $1.4 M_\odot$ neutron star and the maximum mass neutron star, respectively. The left panel corresponds to {Set I}, while the right panel corresponds to {Set II}.}
    \label{fig:eos}
\end{figure}

\section{\label{results} Results}
In this section, we examine the posterior probability distributions for our model that incorporate $\Delta$-resonances. We analyze the relationships between input parameters, isoscalar and isovector components of the nuclear EoS, and key neutron star properties. By evaluating the marginalized probability density functions (PDFs) for nuclear matter (NM), and neutron star observables, we identify correlations and compare trends with existing literature. Our EoS model considers the full baryon octet alongside $\Delta$-resonances, with leptons ($e^-$ and $\mu^-$) ensuring charge neutrality and $\beta-$equilibrium. Using Bayesian parameter estimation, we infer the DDRH model parameters while imposing constraints from nuclear matter saturation properties (e.g., $\rho_0$, $\epsilon_0$, $K_0$, and $J_{\text{sym},0}$), chiral effective field theory ($\chi$EFT) predictions \cite{Drischler_2016}, and astrophysical observations such as PSR J0030+0451 \cite{riley_2019_3386449}, PSR J0740+66 \cite{riley_2021_4697625}, and tidal deformability from GW170817 \cite{PhysRevLett.119.161101}. 
%The density dependence of mesonic couplings follows Eq. \eqref{fx}, reducing the number of free parameters but limiting direct control over density dependence, as seen in models like DDME2 \cite{DDME2} and DD2 \cite{DD2}. Since the couplings in these models are not independent but derived from boundary conditions, they cannot be directly used in Bayesian inference, requiring an iterative approach to self-consistently determine the remaining parameters.

The generated EoSs are required to satisfy causality ($c_s/c < 1$), thermodynamic stability ($dP/d\rho > 0$), the observed maximum neutron star mass ($M_{\text{max}} \geq 2M_\odot$) \cite{2021ApJ...915L..12F}, and a positive symmetry energy at all densities \cite{Abbott_2018}. In the DDRH model, causality ($c_s/c < 1$) is inherently ensured by construction. The saturation density $\rho_0$, appearing in Eq. \eqref{eq:density_dependence}, is determined self-consistently for each model. If $\rho_0$ falls outside a predefined range, the sampler rejects the input parameters, following a method similar to Ref. \cite{Mikhail_2023, Parmar_2024}. Approximately 15,000 valid EoS configurations are generated for both Set I and Set II after evaluating $\sim$150,000 likelihood functions. We use the \textit{corner.py} package \footnote{\url{https://corner.readthedocs.io/en/latest/}} \cite{Foreman-Mackey2016} to visualize the one- and two-dimensional projections of the posterior distributions. In the 2D plots, we display contours at 1$\sigma$ (39.3\%), 68\%, and 90\% confidence intervals (CIs), providing a comprehensive statistical representation of the inferred EoS parameters.

Fig. \ref{fig:par_post} presents the marginalized posterior distributions of the model parameters listed in Table \ref{tab:prior}, considering the constraints defined as Set I and Set II, which differ in the range of symmetry energy. The parameters $g_{\sigma N}$, $g_{\omega N}$, $g_{\rho N}$, $a_{\sigma}$, $a_{\omega}$, $a_{\rho}$, $R_{\sigma \Delta}$, and $R_{\omega \Delta}$, along with the derived quantity $D_{\sigma \omega} = R_{\sigma \Delta} - R_{\omega \Delta}$, are shown with vertical lines indicating their median values. The 68\% confidence intervals (CIs) are displayed at the top of each parameter distribution, while the 2D contour plots illustrate the 1$\sigma$, 2$\sigma$, and 3$\sigma$ confidence intervals. A strong correlation is observed between $g_{\sigma N}$–$g_{\omega N}$ and $a_{\sigma}$–$a_{\omega}$, primarily due to the stringent constraints imposed on the  energy ($\epsilon_0$) and saturation density ($\rho_0$). Set II, which considers a larger symmetry energy ($J$), favors systematically higher values of the model parameters. Notably, both $g_{\sigma N}$ and $g_{\omega N}$ are well-constrained, with posterior distributions favoring the ranges $\approx$ [9.6–9.9] and [11.6–12.0], respectively, compared to their priors. The parameters $R_{\sigma \Delta}$ and $R_{\omega \Delta}$ exhibit values $> 1$, a result similar to Ref. \cite{Marquez2022}. Specifically, $R_{\sigma \Delta}$ is found to be $1.38^{+0.08}_{-0.10}$ for Set I and $1.35^{+0.10}_{-0.16}$ for Set II, while $R_{\omega \Delta}$ is constrained to $1.18^{+0.12}_{-0.11}$ and $1.18^{+0.15}_{-0.11}$, respectively. These values align with prior theoretical expectations, where $R_{\omega \Delta}$ is often assumed to be $\sim 1.10$ \cite{Thapa_2023, Li2019a}. The derived parameter $D_{\sigma \omega}$ is found in the range [0.06–0.29] for Set I and [-0.02–0.28] for Set II, in reasonable agreement with theoretical expectations of $D_{\sigma \omega} \sim [0,0.2]$ \cite{Li2018}, though our results suggest a slightly broader upper bound. The posterior distribution of $D_{\sigma \omega}$ shows a distinct peak for Set I, owing to its narrow symmetry energy range, while Set II exhibits a flatter distribution, indicating greater uncertainty. While $R_{\sigma \Delta}$ and $R_{\omega \Delta}$ do not show strong correlations with other parameters, $D_{\sigma \omega}$ exhibits some level of correlation, suggesting its significance in constraining the model. A more detailed analysis of correlation strengths will be discussed in a later section.

Next, we perform the analysis of nuclear matter properties using the posterior distributions obtained from the computed model parameters. Fig. \ref{fig:nuc_prop} presents the marginalized 1D and 2D posterior distributions of the nuclear matter properties (NMPs) and the $\Delta$ resonance coupling strengths. The compressibility \( K_0 \) lies on the higher end of the currently allowed range, measured as \( 240 \pm 10 \) MeV \cite{PhysRevC.70.024307} or \( 248 \pm 8 \) MeV \cite{PhysRevC.69.041301}. Although Bayesian fitting constrains \( K_0 \) to \( 230 \pm 40 \) MeV, lower values fail to support a 2 \( M_\odot \) star with a $\Delta$-admixed hypernuclear neutron star. Similarly, \( Q_0 \) is also relatively high compared to the posterior analysis of the EoS without $\Delta$ resonances \cite{Malik_2022_1, Mikhail_2023}. 
It was shown in Ref. \cite{Li2019a} that compact star featuring both hyperons and $\Delta$ resonances can be obtained if the value of \( Q_0 \) is large enough using a set of hadronic EoSs derived from relativistic density functional theory. Our results point in the same direction; however, they are obtained by exploring the full parameter space.
Set II estimates a slightly higher incompressibility compared to Set I due to the wider and higher range of \( J \) used in the constraints to account for PREX-II data. The values of \( J \) and \( L \) fall within the range constrained by various experimental analyses \cite{GARG200736, PhysRevC.76.051603, PhysRevLett.94.032701}. Notably, for Set II, \( J \) prefers the lower values within the imposed constraint range, i.e., [33–43] MeV. Therefore, despite the PREX-II constraints, the symmetry energy remains in the lower range, aligning with the current value of \( J = 32.5 \pm 1.8 \) MeV, rather than favoring the higher value estimated by PREX-II.  The slope parameter \( L \) for Set II is slightly larger than that of Set I and does not lie within the region estimated by PREX-II, which suggests \( L = 106 \pm 37 \) MeV.
The lower value of \( L \) in our posterior arises from the use of \(\chi\)EFT predictions to constrain the low-density part of the equation of state. It has been shown that when \(\chi\)EFT constraints are applied at low densities, large values of \( L \), and consequently \( J \), are disfavored \cite{Essick_104, Thakur_2025}. Our results reflect the same trend. Therefore, it seems that the inclusion of $\Delta$ resonances in neuron star EoS does not favour  a high symmetry energy and slope parameter.

Furthermore, using the same RMF formalism as in this work, Bayesian analysis with antikaons was studied in \cite{Parmar_2024}.  The distribution of NMPs appears to show minimal variation between these two studies. 
This trend has been shown by various studies that the posterior distributions of \( J \), \( L \), and other nuclear matter observables remain relatively unchanged across different compositions of matter inside the star, under the currently available constraints on the EoS \cite{Huang:2024rvj}. Since there is no consensus on the magnitude of the $\Delta$ potential in nuclear matter \cite{Li2018}, with values of $U_\Delta/V_N$ typically considered between 1 and 5/3 \cite{Thapa_2023, Li2018, Li2019, Raduta2014}, we have instead varied the coupling strength and computed $U_\Delta/V_N$ as a posterior parameter. We find that the fraction $U_\Delta/V_N$ varies between [1.41-2.18] for Set I and [1.05-2.22] for Set II. It should be noted that in the literature, the appearance of $\Delta$-resonances through a first-order transition is often omitted, despite the fact that they can emerge via a first-order transition, potentially leading to spinodal instabilities \cite{RADUTA2021136070}. Our approach considers both first- and second-order transitions, favouring a slightly deeper potential for $U_\Delta$.  The coupling strengths $R_{\sigma \Delta}$ and $R_{\omega \Delta}$ do not exhibit a strong correlation with the nuclear matter properties, whereas their difference, $D_{\sigma \omega}$, shows a relatively strong correlation.

In Fig. \ref{fig:tov_post}, we present the  marginalized posterior distributions of NS properties, including $M_{\text{max}}$,  central pressure ($P_c$) in MeV-fm$^{-3}$, radius at maximum mass ($R_{\text{max}}$) in km, tidal deformability at maximum mass ($\Lambda_{\text{max}}$), radius of the canonical $1.4 M_\odot$ star ($R_{1.4}$) in km, and its tidal deformability ($\Lambda_{1.4}$), $f_{1.4}$ and $p1_{1.4}$ mode frequency in kHz and their respective decay time in units of seconds are shown along with the distribution of  $R_{\sigma \Delta}$, $R_{\omega \Delta}$ and  $D_{\sigma \omega}$. The 2D ellipses and vertical lines have the same interpretation as in Fig. \ref{fig:par_post}.  $M_{\text{max}}$ of the neutron star remains consistently above $2 M_\odot$ due to the constraints imposed in our model. Notably, Set II exhibits a broader mass distribution owing to the wide range of symmetry energy values considered. $M_{\text{max}}$ is relatively lower compared to many studies involving purely nucleonic or hyperonic matter, primarily due to the appearance of $\Delta$ resonances.
 The predicted radius of a $1.4 M_\odot$ neutron star ($R_{1.4}$) falls comfortably within the recently reported range of  $R_{1.4} = 12.01^{+0.78}_{-0.77}$ km \cite{Huth2022}, which investigated constraints on neutron-star matter through microscopic and macroscopic collisions, utilizing data from $\chi$EFT, multi-messenger astrophysics, and heavy-ion collision (HIC) experiments. Compared to purely nucleonic \cite{Malik_2022} or hyperonic equations of state (EoS) \cite{Malik_2022_1}, our results for $R_{1.4}$ are lower, highlighting the impact of $\Delta$ baryons in reducing the neutron star radius making it more compact.  The dimensionless tidal deformability $\Lambda_{1.4}$ falls well within the range allowed by NICER \cite{Miller_2021} and the GW170817 event \cite{PhysRevLett.119.161101}. Additionally, we compute the fundamental ($f$-mode) and  ($p1$-mode) oscillation frequencies using a fully general relativistic formalism to avoid the limitation of Cowling approximation \cite{guha2023}. 
 Our reported values for the $f$-mode frequency are: $1.97^{+0.17}_{-0.22}$ kHz for Set I and $1.77^{+0.26}_{-0.14}$ kHz on 68\% CI. These values align well with gravitational wave constraints from GW170817. The 90\% credible interval for the $f$-mode frequency in GW170817 was reported as $1.43 \text{ kHz} < f < 2.90 \text{ kHz}$ for the more massive component and $1.48 \text{ kHz} < f < 3.18 \text{ kHz}$ for the less massive component \cite{Pratten2020}. The $f$-mode oscillation frequency plays a crucial role in neutron star astrophysics as it is directly linked to the star's compactness and EoS. It is expected to be a key observable in future gravitational wave detections. The $p1$-mode, on the other hand, provides insight into the high-frequency oscillations driven by pressure waves, which can help probe the inner structure of neutron stars. Unlike bulk nuclear matter properties, the parameter $D_{\sigma \omega}$ strongly influences neutron star properties in $\Delta$-admixed hyperonic neutron stars. We find a strong correlation between key neutron star observables such as $R_{1.4}$, $\Lambda_{1.4}$, and $f_{1.4}$ and  $D_{\sigma \omega}$, emphasizing the significant role of $\Delta$ baryons in neutron star structure and oscillation dynamics.   

\begin{figure}%[h!]
    \centering
    \includegraphics[scale=0.34]{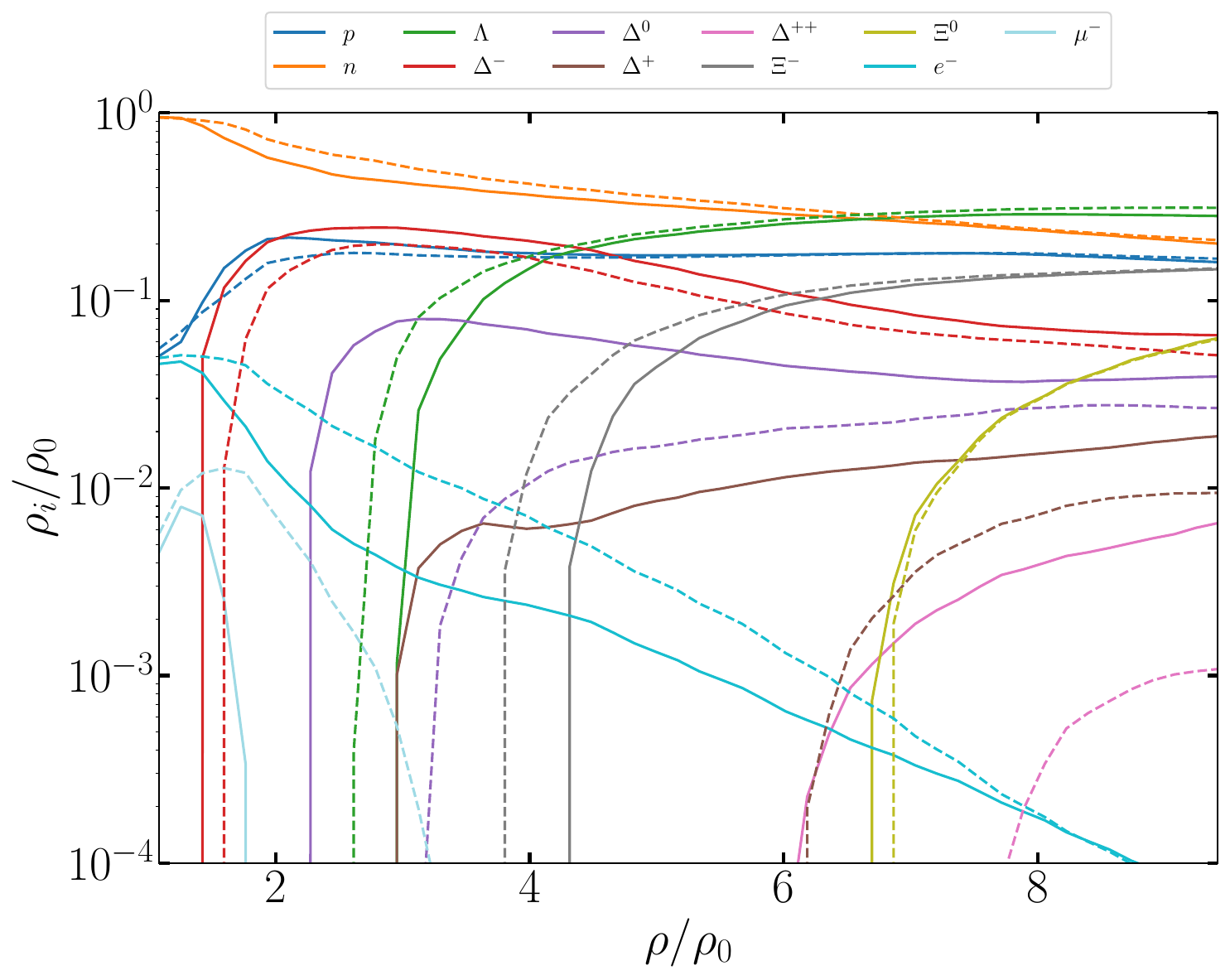}
\caption{Particle fractions of various species inside the neutron star for Set I (solid lines) and Set II (dashed lines). The fractions correspond to the median values at each baryon density, representing the average composition.}
    \label{fig:pf}
\end{figure}

\begin{figure*}
    \centering
    \includegraphics[scale=0.35]{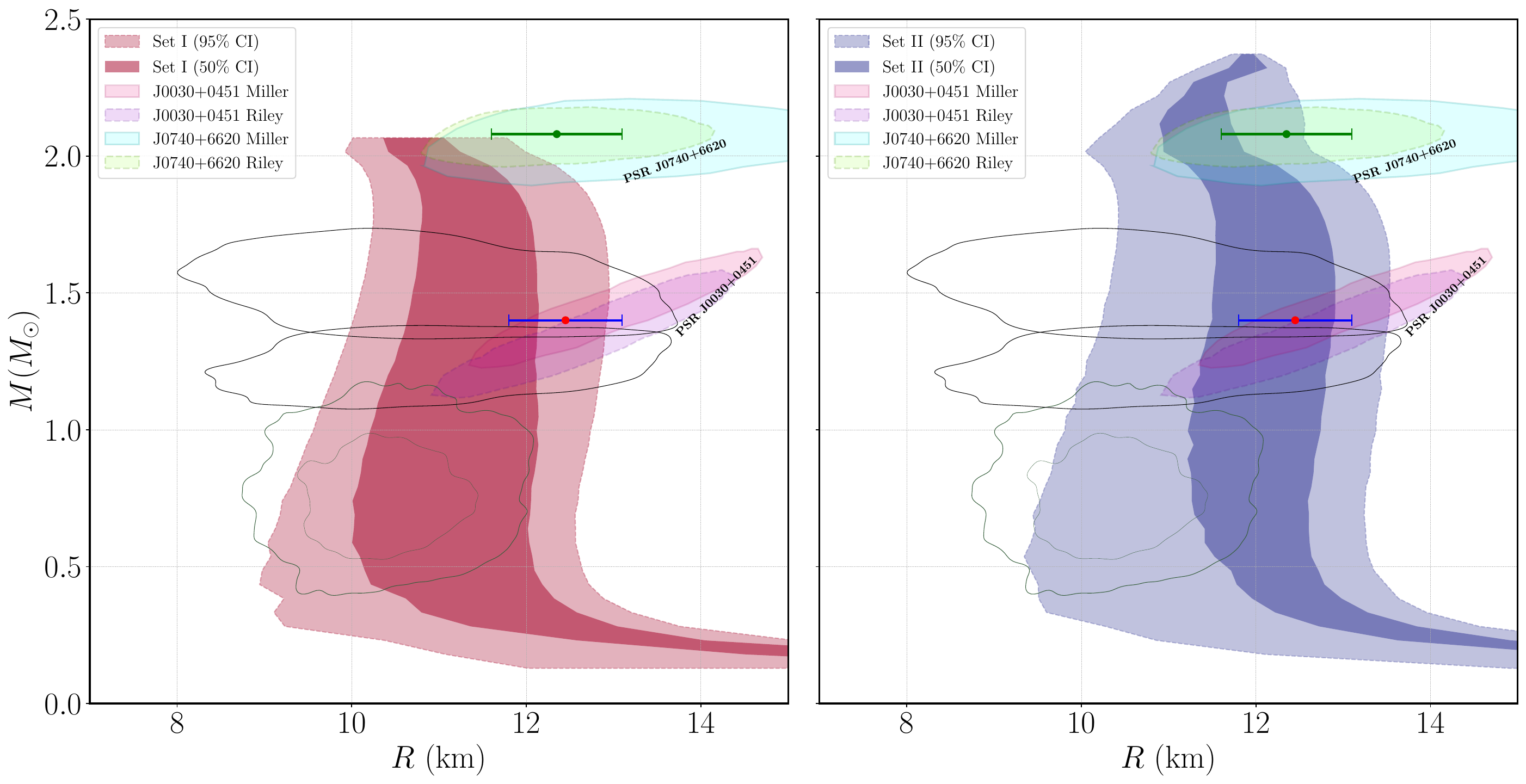}
\caption{The  M-R relation for the ensemble of  EoSs in Fig. \ref{fig:eos}. The M-R posterior is compared with the 90\% CI for the binary components of the GW170817 event, represented by the black lines. The double-headed blue and green lines represent the radius constraints by Miller \textit{et al.} \cite{Miller_2021_1.4} for neutron stars of masses 1.4 and 2.08 M$_\odot$, respectively. The shaded regions in dark and light blue represent the constraints from the heavy mass pulsar PSR J0740+6620 by Riley \textit{et al.} \cite{Riley_2019} and Miller \textit{et al.} \cite{Miller_2019}, respectively. Similarly, the shaded regions in pink and orange represent the constraints from the millisecond pulsar PSR J0030+0451 by Riley et al. \cite{riley_2021_4697625} and Miller \textit{et al.}. The solid Green line corresponds to the  50 \% and 90 \% CI for supernova remnant HESS J1731-347 \cite{Doroshenko2022}. }
    \label{fig:mr}
\end{figure*}

Fig. \ref{fig:eos} illustrates the EoS for neutron star matter with $\Delta$-resonance admixed hyperonic components. 
The EoS is compared with recent estimations derived from a combination of microscopic nuclear theory and multi-messenger astrophysics constraints \cite{Huth2022}, as well as the EoS constraints from GW170817.  Set II exhibits a stiffer EoS band compared to Set I, primarily due to its larger symmetry energy and incompressibility. 
This study explores the emergence of $\Delta$ resonances through both first- and second-order phase transitions. Notably, the first-order transition introduces spinodal instabilities, a feature often overlooked in prior investigations of $\Delta$ resonances. Our analysis reveals that at lower densities, the 50$\%$ and 95$\%$ CI bands display a broader spread, attributed to the early onset of $\Delta$ resonances. Specific values of the coupling parameters $R_{\sigma \Delta}$ and $R_{\omega \Delta}$ trigger spinodal instabilities, which we mitigate using a Maxwell construction, following the method outlined in \cite{RADUTA2021136070}. This approach leads to a pronounced softening of the EoS at low densities. Furthermore, the occurrence of a first-order phase transition at such densities suggests the possibility of a twin-star configuration, akin to the scenario observed in hadron-quark transitions \cite{universe6060081}.
Both Set I and Set II EoS results are consistent with the EoS constraints derived from GW170817. It should be noted that although the symmetry energy in Set II was constrained using values derived from the PREX-II experiment, the inclusion of $\Delta$ resonances disfavors an excessively high symmetry energy and slope parameter, as discussed in previous sections. This tension between the PREX-II results and the current understanding of neutron star properties has been reported in the literature \cite{Reed2021} and persists even with the inclusion of $\Delta$ resonances.
 Additionally, we find that a higher value of the parameter $D_{\sigma \omega}$ leads to further softening of the EoS at low and intermediate densities. This highlights the crucial role of $\Delta$ resonances in shaping the neutron star matter EoS, particularly at lower densities where their effects become more pronounced.

To highlight the role of $\Delta$ resonances in the EoS, Fig. \ref{fig:pf} shows the particle fractions of various species inside the neutron star. The solid lines correspond to Set I, while the dashed lines represent Set II. The particle fractions at each density correspond to median values, providing an average composition profile. The $\Delta^-$ resonance appears first at low densities, followed by $\Delta^0$ and then the $\Lambda$ hyperon. The $\Delta^+$ emerges at moderate densities, while the $\Delta^{++}$ appears only at very high densities with minimal contribution. 
The $\Sigma$ hyperon is absent owing to the repulsive interaction in nuclear matter, while the $\Xi^-$  appears at intermediate densities and the $\Xi^0$ at higher densities. The early onset of $\Delta^-$ and $\Delta^0$ at low densities, leads to a decrease in the muon fraction due to charge conservation. At lower densities, $\Delta^-$ and $\Delta^0$ play a dominant role, whereas at higher densities, the $\Lambda$ hyperon becomes more significant. This results in a softer EoS at low densities and a stiffer EoS at high densities. A higher symmetry energy in Set II results in a delayed onset of $\Delta$ resonances compared to Set I, which significantly impacts the overall EoS and the particle composition of dense matter. Moreover, in Set II, the appearance of $\Lambda$ and $\Xi$ hyperons occurs at lower densities than in Set I.

\begin{figure*}
    \centering
    \includegraphics[scale=0.35]{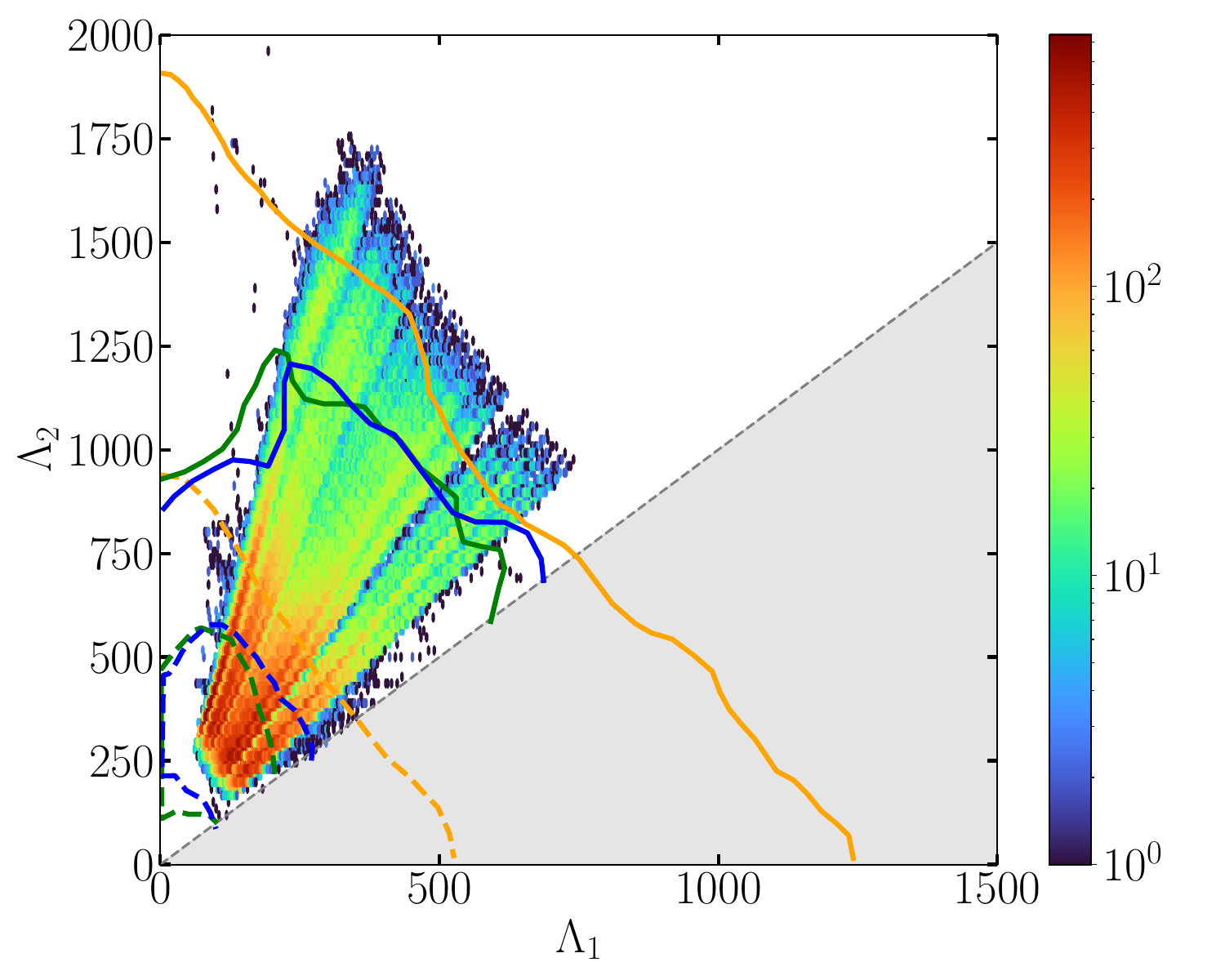}
    \includegraphics[scale=0.35]{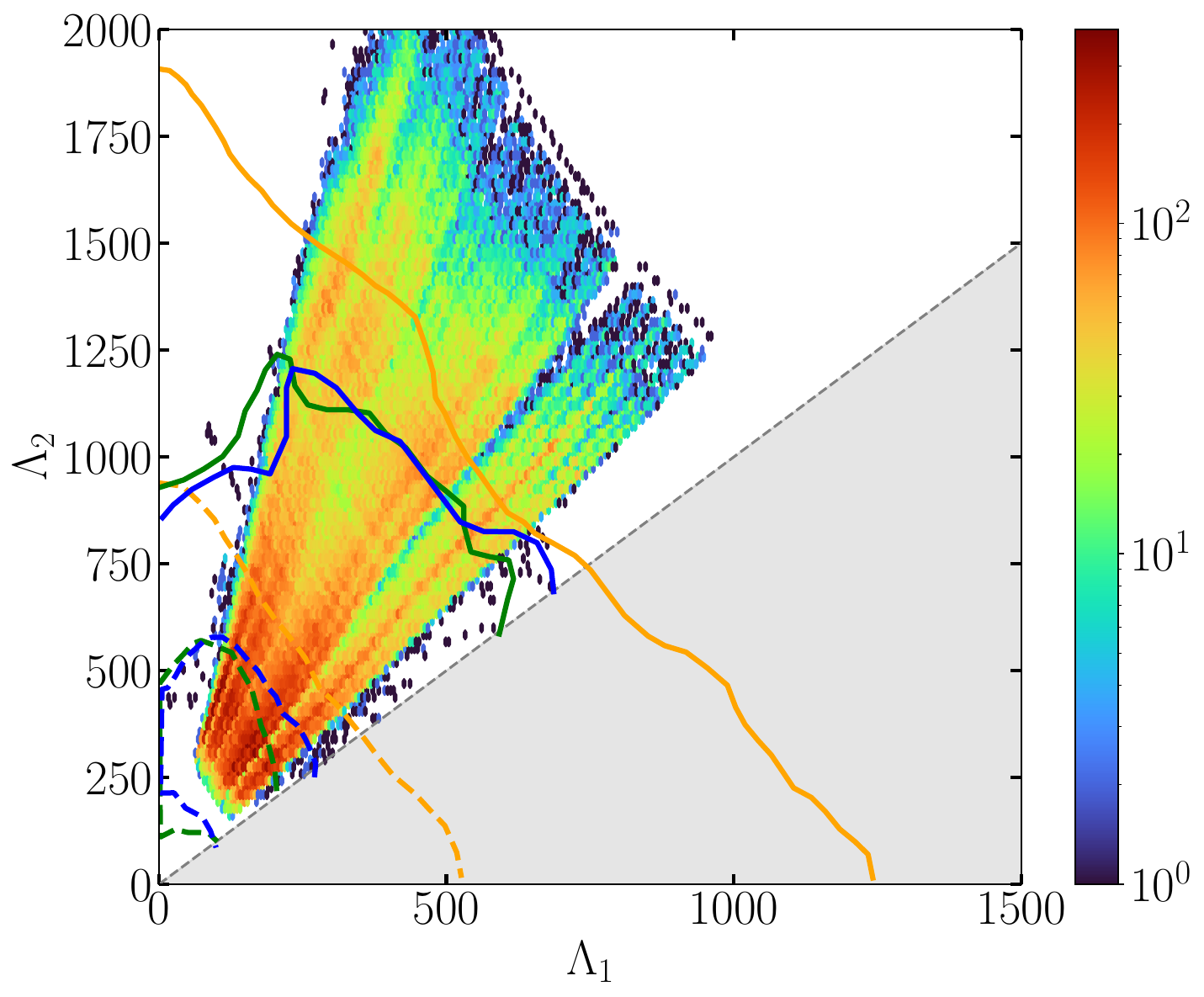}
\caption{The posterior probability distributions for $P(\Lambda_1, \Lambda_2)$ for Set I (left) and Set II (right) respectively, are illustrated. Here, $\Lambda_1$ and $\Lambda_2$ refer to the dimensionless tidal deformability parameters associated with the BNS merger observed in the GW170817 event. The figure showcases the unphysical region where $\Lambda_1$ is less than $\Lambda_2$, illustrated in gray shading. Meanwhile, the green, blue, and orange lines signify the 50\% (dashed) and 90\% (solid) credible levels for the posterior distributions obtained using EoS-insensitive relations, a parametrized EoS without a maximum mass requirement, and independent EoSs, respectively, from the GW170817 event \cite{Abbott_2018}. the colorbar represents the probability on log scale.}
    \label{fig:lam1lam3}
\end{figure*}

Fig. \ref{fig:mr} presents the 95\% CI bands for the mass-radius (M-R) relationship derived from the ensemble of EoSs obtained in Fig. \ref{fig:eos}. The M-R relation is compared with various astrophysical constraints, including those from the millisecond pulsars PSR J0030+0451 and PSR J0740+6620, the binary components of the GW170817 event, and the radius constraints reported by Miller \textit{et al.} \cite{Miller_2021_1.4} for neutron stars with masses of $1.4 M_\odot$ and $2.08 M_\odot$.  Furthermore, we compared our results with the recently measured compact low-mass object in the supernova remnant HESS J1731-347, which has an estimated mass of $M = 0.77^{+0.20}_{-0.17} M_\odot$ and a radius of $R = 10.4^{+0.86}_{-0.78}$ km \cite{Doroshenko2022}. The M-R posterior agrees well with the observational constraints of PSR J0030 + 0451, PSR J0740 + 6620, and the binary components of the GW170817 event. 
A notable consequence of $\Delta$ resonance formation is a systematic reduction in stellar radii, which serves as a distinctive feature of our results. 
One of the most compelling aspects of our findings is the remarkable agreement between the M-R posterior and the recently detected low-mass compact object in the supernova remnant HESS J1731-347. 
Although this object was not initially regarded as a constraint in neutron star modeling, our results naturally accommodate its observed mass and radius. 
This is particularly striking, as many previous studies have encountered difficulties in explaining such a low-mass compact star within the framework of conventional EoSs. 
Instead, alternative interpretations have been proposed, classifying it as a neutron star with heavy baryons \cite{LI2023138062}, antikaon condensation \cite{VESELSKY2025139185},  strange star \cite{Sagun_2023, Horvath_2023}, or even a compact star with a dark matter component \cite{Kubis2023}.
A major challenge is reconciling the need for a low-radius solution at lower masses, as required by HESS J1731-347, with the requirement for a maximum mass exceeding $2 M_\odot$, as imposed by GW170817 and pulsar data. This necessitates an EoS that is soft at low densities while becoming sufficiently stiff at high densities to support massive neutron stars. The inclusion of $\Delta$ resonances naturally achieves this balance. In the literature, exotic scenarios such as twin-star branches from first-order hadron-quark phase transitions and deep antikaon potentials have been explored to explain this dichotomy \cite{LI2023138062, VESELSKY2025139185, Sagun_2023, Kubis2023}.  Our results demonstrate that the inclusion of $\Delta$ resonances provides a natural resolution: their appearance in the outer layers of the core softens the EoS sufficiently to accommodate the compact nature of HESS J1731-347 while still supporting massive neutron stars exceeding $2 M_\odot$, as required by NICER and GW observations. Thus, the presence of $\Delta$ resonances offers a unified explanation that simultaneously satisfies both the high-mass constraint from GW and pulsar observations and the low-mass constraint from HESS J1731-347, reinforcing their critical role in neutron star structure.

To further understand the effect of $\Delta$ resonances, in Fig. \ref{fig:lam1lam3} we present the probability distribution $P(\Lambda_1, \Lambda_2)$, where $\Lambda_1$ and $\Lambda_2$ represent the dimensionless tidal deformability parameters corresponding to the BNS merger observed in the GW170817 event.  To compute $P(\Lambda_1, \Lambda_2)$, we adopt a chirp mass of $\mathcal{M} = 1.188 M_\odot$ and a mass ratio defined as $q = m_2/m_1$, constrained within the range $0.7 < q < 1$. We fix $m_1$ and determine $m_2$ such that the conditions $\mathcal{M} = 1.188 M_\odot$, $0.7 < q < 1$, and $2.73 \leq m_1 + m_2 \leq 2.78 M_\odot$ are satisfied. Using the computed values of $m_1$ and $m_2$, we then evaluate $\Lambda_1$ and $\Lambda_2$. The results are compared with posterior distributions derived using EoS-insensitive relations, a parameterized EoS without a maximum mass constraint, and independent EoS models inferred from the GW170817 event \cite{Abbott_2018}.  Our posterior results show excellent agreement with those derived from the GW170817 event, with the region of highest probability falling within the 50\% credible interval of the GW170817 BNS merger. Although Set II exhibits a broader distribution compared to Set I, both calculations strongly suggest that the binary in GW170817 favors the presence of $\Delta$-isobars in addition to hypernuclear matter. This finding supports the argument of Ref.~\cite{Li2019a}, which posits that delta resonances with an attractive $\Delta$ potential in nuclear matter can explain GW170817 in a manner similar to a strong hadron-quark phase transition. However, unlike the hadron-quark phase transition, $\Delta$ resonances appear at relatively lower densities, potentially providing a unique observational signature to distinguish the two scenarios. One can analyze these in  the plot of the two tidal deformabilities $\Lambda_1$ and $\Lambda_2$ of BNSs \cite{Christian_2019}, as well as in the weighted average tidal deformability $\bar{\Lambda}$ at a given chirp mass $\mathcal{M}$ \cite{Sophia_2019}, and in General-Relativistic Neutron-Star Mergers \cite{Elias_2019}.

\begin{figure}
    \centering
    \includegraphics[width=1\linewidth]{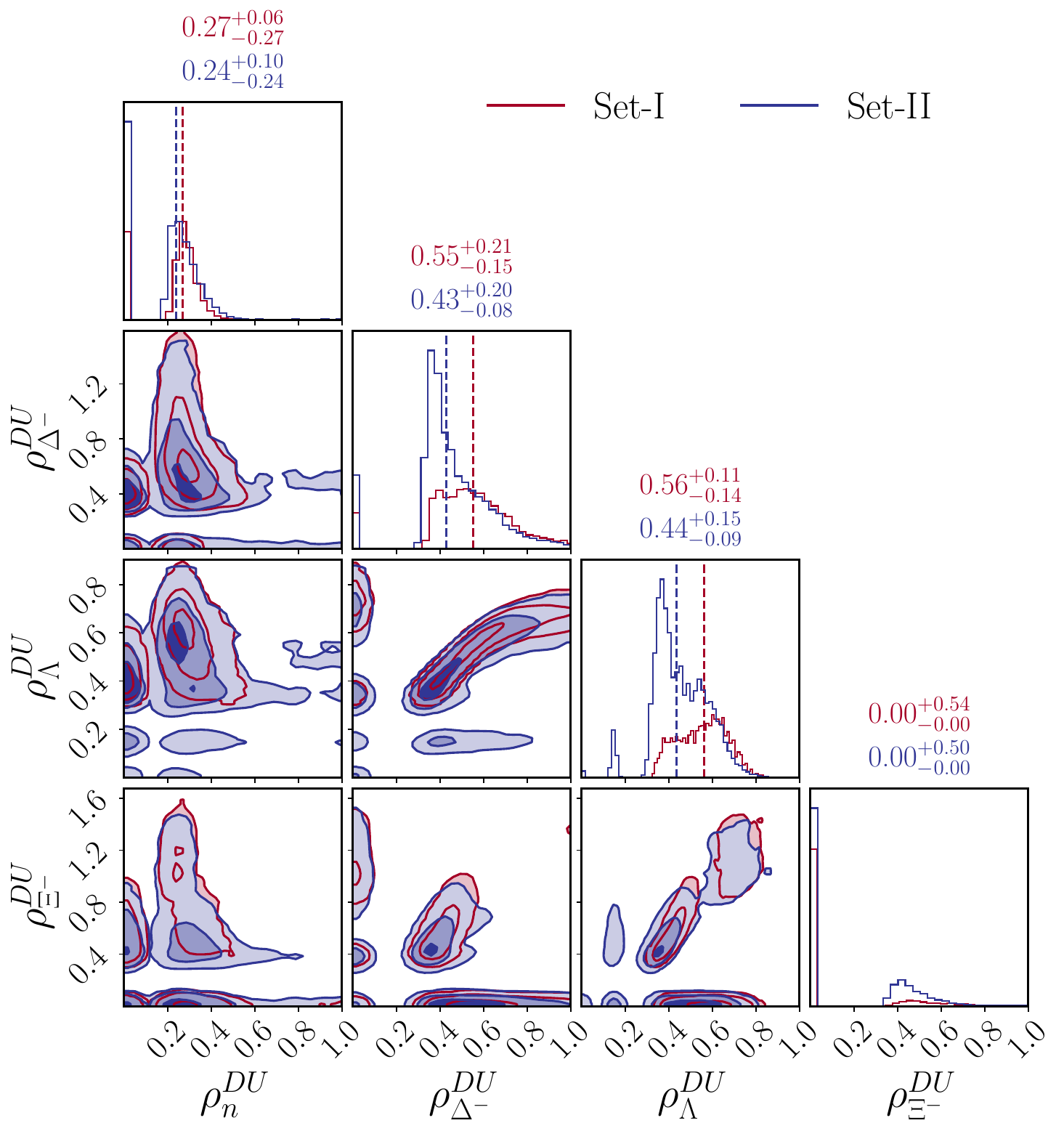}
    \caption{ The marginalized posterior distribution of the direct Urca (DU) processes:  
$n \rightarrow p + e^- + \bar{\nu}_e$,  
$\Delta^- \rightarrow \Lambda + e^- + \bar{\nu}_e$,  
$\Lambda \rightarrow p + e^- + \bar{\nu}_e$, and  
$\Xi^- \rightarrow \Lambda + e^- + \bar{\nu}_e$.
}
    \label{fig:du}
\end{figure}

In neutron stars, the direct Urca (DU) process is a crucial mechanism that enables rapid cooling via neutrino emission when the proton fraction exceeds a critical threshold. The inclusion of hyperons lowers this threshold, allowing the DU process to occur at lower densities, thereby significantly enhancing cooling. In Fig. \ref{fig:du}, we show the DU process threshold in a $\Delta$-admixed hyperonic neutron star. We consider four key DU processes: $n \rightarrow p + e^- + \bar{\nu}_e$, $\Delta^- \rightarrow \Lambda + e^- + \bar{\nu}_e$, $\Lambda \rightarrow p + e^- + \bar{\nu}_e$, and $\Xi^- \rightarrow \Lambda + e^- + \bar{\nu}_e$.  For a general DU reaction, $B_1 \rightarrow B_2 + l + \bar{\nu}_l$ and its inverse process $B_2 + l \rightarrow B_1 + \nu_l$, where $B_1$ and $B_2$ represent the participating baryons, $l$ is the lepton (electron or muon), and $\nu_l$ ($\bar{\nu}_l$) denotes the corresponding neutrino (antineutrino).  For the DU process to occur, the phase-space momentum conservation condition must be satisfied, imposing a constraint on the Fermi momenta of the participating particles: $|p_F^{B_2} - p_F^l| \leq p_F^{B_1} \leq p_F^{B_2} + p_F^l$ \cite{Haensel_1994}, where $p_F^{B_1}$, $p_F^{B_2}$, and $p_F^l$ are the Fermi momenta of the initial baryon, final baryon, and lepton, respectively. If this condition is not met, the process is kinematically forbidden due to momentum conservation. With the inclusion of $\Delta$ resonances, the proton fraction increases while the electron fraction decreases, leading to a lower threshold for the neutron direct Urca (DU) process. For the neutron DU process, a sharp peak at zero density indicates that the process is not allowed in those cases. However, when allowed, the threshold density $\rho_n^{DU}$ typically occurs around twice the nuclear saturation density. The threshold densities for the $\Delta^-$ and $\Lambda$ DU processes are nearly identical. These densities fall within the mass range of a $1.4 M_\odot$ neutron star and are favored in most EoSs within the posterior distribution. The $\Xi^-$ DU process is absent due to the very low lepton fraction at high densities, as seen in Fig. \ref{fig:pf}. A strong positive correlation is observed between $\rho_{\Delta^-}^{DU}$ and $\rho_{\Lambda}^{DU}$, indicating a close connection between their onset conditions. The possible existence of neutron, $\Delta^-$, and $\Lambda$ DU processes within a $1.4 M_\odot$ neutron star suggests that $\Delta$ resonances could significantly modify the neutrino emission rates, particularly when $\Delta$ baryons populate the outer layers of the neutron star core.

\begin{figure*}
    \centering
    \includegraphics[width=\textwidth]{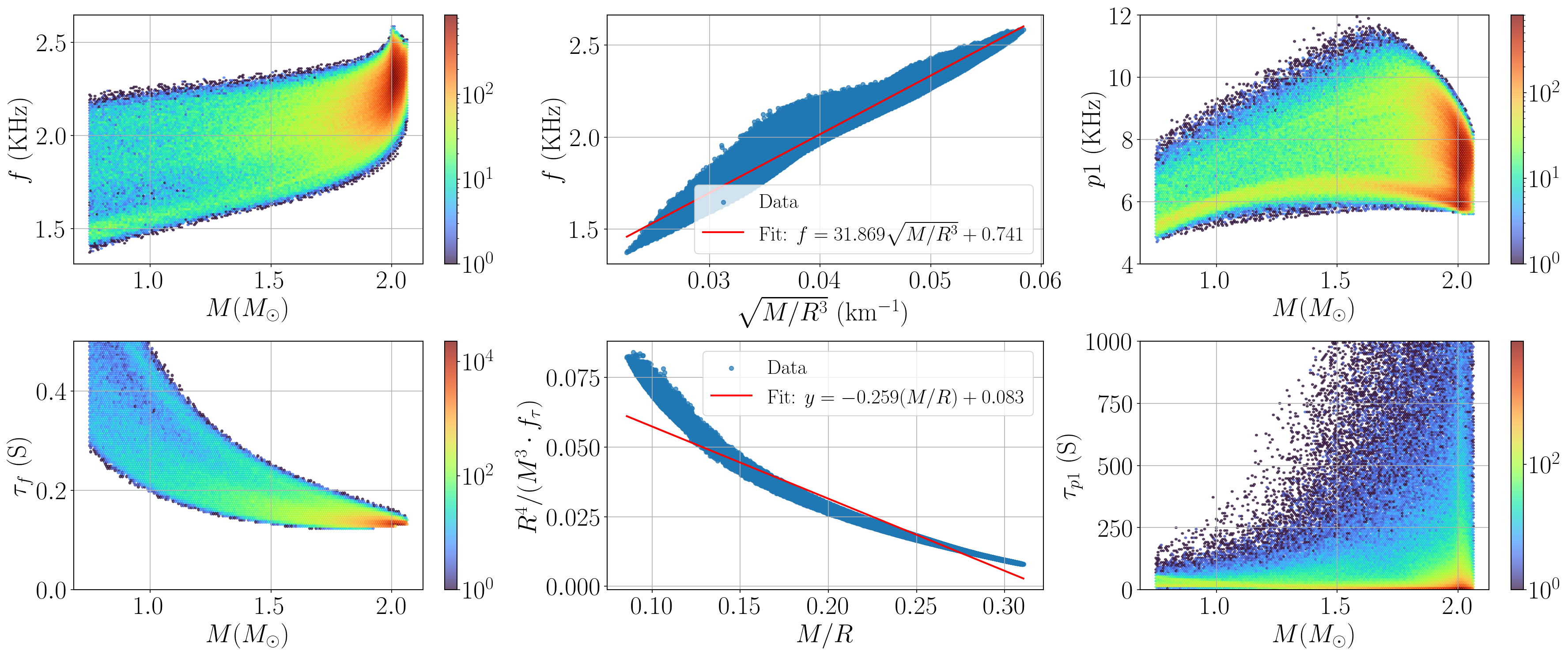}
  \caption{
    \textbf{Left panel:} The posterior probability distribution of $f$-mode oscillation frequency and the corresponding decay time as a function of NS mass. The colorbar represents the probability density on a logarithmic scale.  
    \textbf{Middle panel:} The empirical relation between the $f$-mode frequency and the average stellar density, as well as the relation between the dimensionless gravitational-wave (GW) damping time and the stellar compactness. The red lines indicate the linear fit of the form $y = ax + b$.
    \textbf{Right panel:} The posterior probability distribution of $p1$-mode oscillation frequency and the corresponding decay time as a function of NS mass.
    } 
    \label{fig:fm}
\end{figure*}

Gravitational wave asteroseismology has emerged as a powerful tool for probing neutron star interiors, offering crucial insights into their composition and EoS \cite{sotani2021}. Among the various oscillation modes, the fundamental $f$-mode is particularly significant, as it is primarily governed by the star’s bulk properties, including mass, radius, and compactness. The frequency and damping time of the $f$-mode are directly linked to the EoS, making them key observables for constraining dense matter physics. Future GW observatories, such as the Einstein Telescope and LISA, are expected to enable high-precision measurements of gravitational waves generated from various sources, providing deeper constraints on neutron star structure and the possible existence of exotic matter at supranuclear densities.  In Fig. \ref{fig:fm}, we present the posterior probability distribution of the $f$-mode oscillation frequency and the corresponding damping time as a function of neutron star mass within the general relativity framework. We use the ensemble of EoSs from Set I, as it is based on more stringent constraints on $J$ compared to Set II, which is informed by PREX II. As shown earlier, our model with $\Delta$-resonances does not favor the higher values of $J$ and $L$ prescribed by PREX II. The $f$-mode frequency ranges from 1.5 to 2.6 kHz,  increasing monotonically with NS mass. At $1.4 M_{\odot}$, the most probable value is found to be $1.97^{+0.17}_{-0.22}$ kHz at 68\% CI.  In Table \ref{tab:fmode_frequencies}, we compare our predicted $f_{1.4}$ value for delta-admixed hypernuclear matter with results from previous studies using different EoS with various compositions. 
It can be seen that the $f$-mode frequency is relatively higher for the EoS with $\Delta$-admixed hyperons due to the increased compactness of the star in the presence of $\Delta$-resonances. Moreover, as stated earlier, our values are consistent with those obtained from GW170817 \cite{Pratten2020}. The damping time decreases monotonically with mass, and the obtained ranges for both $f$-mode frequency and damping time are consistent with literature results for purely nucleonic as well as hyperonic EoS.

\begin{table}[ht]
    \centering
    \caption{Reported $f$-mode frequencies and damping times for neutron stars with a mass of 1.4 M$_{\odot}$.}
    \label{tab:fmode_frequencies}
    \resizebox{\linewidth}{!}{ % Fit table within page width
    \begin{tabular}{l c c c} % Corrected column formatting
        \toprule
        \textbf{Study} & \textbf{$f_{1.4}$ (kHz)} & \textbf{$\tau$ (s)} & \textbf{Framework} \\
        \midrule
        \textbf{This work} & $1.97^{+0.17}_{-0.22}$ & $0.19^{+0.05}_{-0.03}$ & FGR \\
        Guha Roy et al. (2023) \cite{guha2023} & $1.8^{+0.7}_{-0.12}$ & $0.2 – 0.25$ & FGR \\
        Wen et al. (2019) \cite{wen2019} & 1.67 – 2.18 & 0.155 – 0.255 & FGR \\
        %Pradhan et al. (2023) \cite{Pradhan_2022} & 1.47 – 2.45 & 0.13 – 0.51 & GR \\
        Mohanty et al. (2024) \cite{2024arXiv241016689R} & $1.75^{+0.23}_{-0.15}$ & $-$ & FGR \\
        \bottomrule
    \end{tabular}}
\end{table}

In the literature, empirical relations are often used to quantify the relationship between the $f$-mode frequency, stellar density, and compactness. We use the following two empirical relations forms \cite{Andersson_1998}:

\begin{equation}
    f = a \sqrt{\frac{M}{R^3}} + b
\end{equation}
and 
\begin{equation}
\label{eq:tau}
    \frac{R^4}{M^3 \tau_f} = a \frac{M}{R} + b.
\end{equation}

These empirical relations are shown in the middle panel of Fig. \ref{fig:fm}, along with the fitted values of $a$ and $b$. When compared with results from other studies using nucleonic or hyperonic EoSs \cite{Pratten2020, guha2023, Kumar_2023}, our values, while comparable, exhibit variations across different works, highlighting their model dependencies. Furthermore, empirical relation \eqref{eq:tau} deviates for small values of $M/R$. In the presence of Delta resonances, the EoS becomes significantly softer at low densities, leading to a breakdown of the empirical relation. Additionally, this relation is known to be invalid for $M/R > 0.25$ \cite{Tsui_2005}. It can be observed that the linear relation in \eqref{eq:tau} does not provide a good fit. Instead, a second-order polynomial offers a better explanation, as demonstrated in \cite{Tsui_2005}. In the right panel of Fig.~\ref{fig:fm}, we present the distribution of the $p1$ mode frequency. The $p1$-modes  exist for both radial and non-radial oscillations, with their frequencies determined by the time it takes for acoustic waves to propagate across the star. The $p1$-mode frequency varies between 4 and 11 kHz, increasing with mass up to a certain point before subsequently decreasing. Unlike $\tau_f$, the damping time $\tau_p1$ exhibits a wider spread, increasing with mass and reaching values on the order of a few hundred seconds. For the $\Delta$ admixed neutron star we found $p1_{1.4}= 7.47^{+1.35}_{-1.10}$ kHz

\begin{figure*}
    \centering
    \includegraphics[scale=.7]{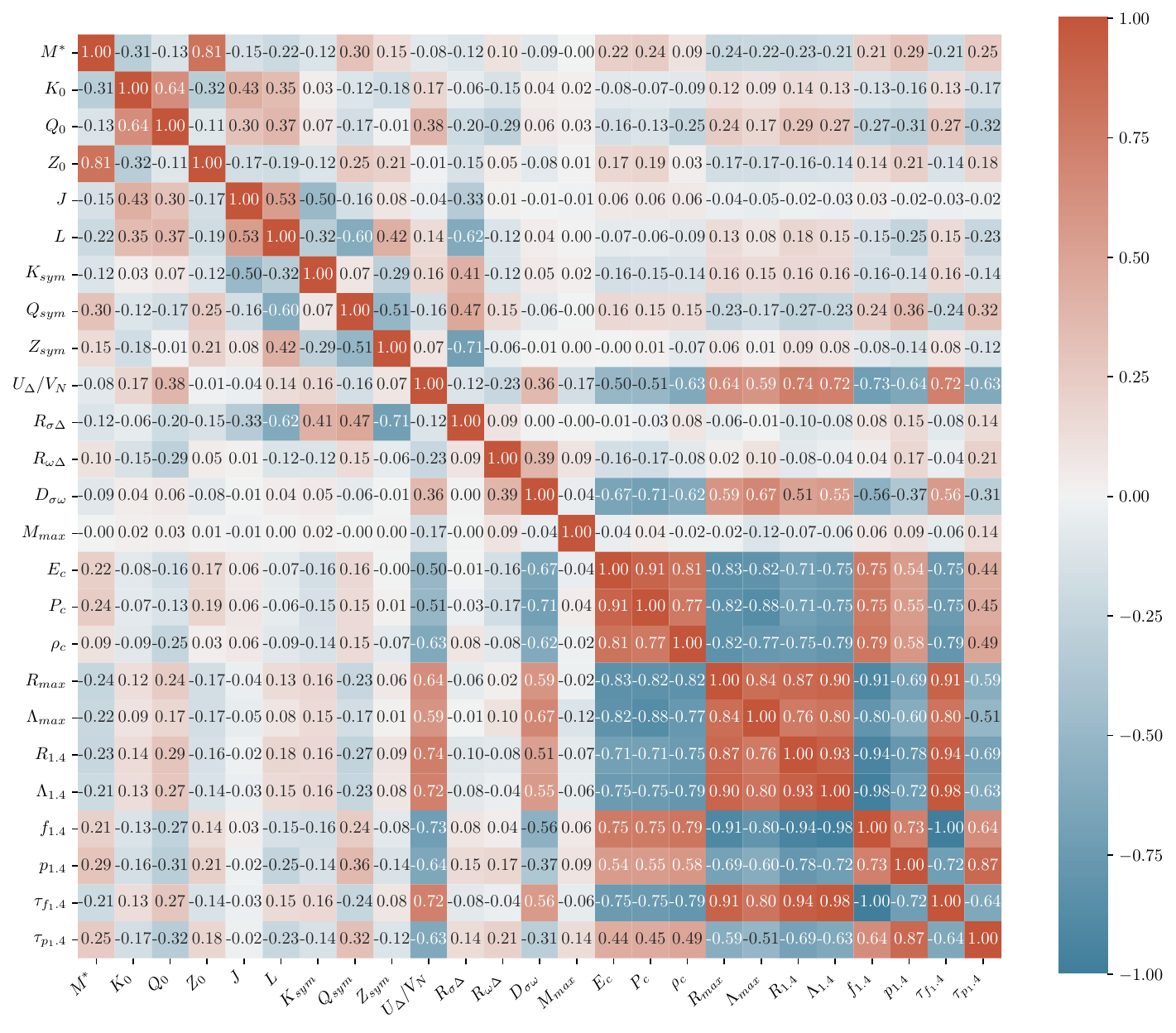}
    \caption{Kendall rank correlation matrix among various nuclear matter properties at the saturation density (i.e.  energy ($e_0$), incompressibility coefficient $K_0$, the skewness $Q_0$, and the kurtosis $Z_0$, symmetry energy coefficient $J$, the slope $L$, the curvature $K_{sym}$, the skewness $Q_{sym}$, (and the kurtosis $Z_{sym}$) and effective mass ($m^*$), $R_{\sigma \Delta}$, $R_{\omega \Delta}$, $D_{\sigma \omega}$  and the neutron star properties with  maximum mass ($M_{max}$), central energy, pressure and density $E_c$, $P_c$ and  $\rho_c$ respectively, radius corresponding to maximum mass ($R_{max}$), tidal defomability corresponding to maximum mass ($\Lambda_{max}$), and the radius and tidal deformability for neutron star mass 1.4 M$_\odot$ $f_{1.4}$ and $p_{1.4}$ mode frequency in $kHz$ and their respective decay time in Sec. The color bar represents the strength of the correlation, with blue representing a strong negative and orange representing a strong positive correlation. The value in white text represent strong correlations.}
    \label{fig:corr}
\end{figure*}

\begin{figure*}
    \centering
    \includegraphics[scale=0.28]{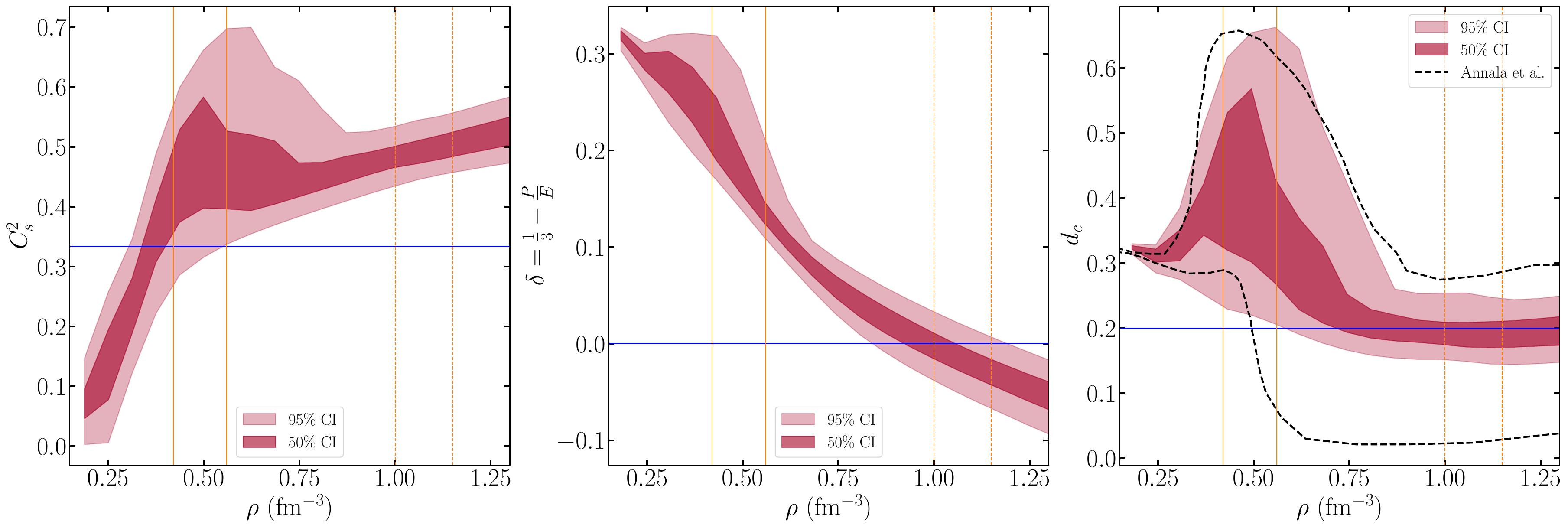}
    \caption{The speed of sound squared (\(C_s^2\)), trace anomaly (\(\delta\)), and conformality factor (\(d_c\)) for the ensemble of EoS in our posterior with the \(\Delta\) resonance admixed hypernuclear matter. The vertical solid and dashed lines represent the central densities of a \(1.4M_{\odot}\) neutron star and the maximum mass neutron star, respectively. The black dashed line represents the 90\% credible interval (CI) value of \(d_c\) derived from Annala et al. \cite{Annala2023}}
    \label{fig:cs}
\end{figure*}

We now discuss the general correlation among various nuclear matter properties, neutron star parameters, and delta resonance coupling parameters, complimenting our analysis in Figs. \ref{fig:par_post}, \ref{fig:nuc_prop} and  \ref{fig:tov_post}.  In Fig. \ref{fig:corr} we present the Kendall rank correlation matrix among different nuclear matter properties at saturation density, including effective mass (\(m^*\)), incompressibility coefficient (\(K_0\)), skewness (\(Q_0\)), kurtosis (\(Z_0\)), symmetry energy coefficient (\(J\)), slope parameter (\(L\)), curvature (\(K_{\text{sym}}\)), skewness (\(Q_{\text{sym}}\)), and kurtosis (\(Z_{\text{sym}}\)). Additionally, we consider the  delta resonance coupling parameters (\(R_{\sigma \Delta}\), \(R_{\omega \Delta}\), \(D_{\sigma \omega}\),\(U_{\Delta}/V_N\) ), and neutron star properties such as maximum mass (\(M_{\text{max}}\)), central energy density (\(E_c\)), central pressure (\(P_c\)), and central density (\(\rho_c\)).  Furthermore, we include the radius corresponding to the maximum mass (\(R_{\text{max}}\)), tidal deformability at maximum mass (\(\Lambda_{\text{max}}\)), and the radius and tidal deformability for neutron stars of mass 1.4 \(M_\odot\). We also examine the fundamental (\(f_{1.4}\)) and pressure (\(p_{1.4}\)) mode frequencies, along with their respective decay times. The color bar in the correlation matrix represents the strength of the correlation, where blue indicates a strong negative correlation, and orange signifies a strong positive correlation. The values in white text highlight strong correlations.

Several studies in the literature have investigated the correlations between nuclear matter properties at saturation density, astrophysical observables, and their linear combinations \cite{Parmar_2024, Alam_2016, Margueron_2018, Malik_2018, Weissenborn2012, Hornick_2018, Tsang_2012, Mikhail_2023, PhysRevD.107.103054, Parmar_3, guha2023}. In our study, It has been observed that there is no significant correlation between isoscalar and isovector parameters, as also reported in \cite{Parmar_2024, Mikhail_2023}. A strong positive correlation is found between the effective mass (\(m^*\)) and the kurtosis parameter (\(Z_0\)), consistent with the findings of \cite{Mikhail_2023, Parmar_2024}, which explored the dense matter EoS using a similar formalism and Bayesian approach, albeit with a different composition. We also observe a negative correlation between the slope parameter (\(L\)) and the skewness of the symmetry energy (\(Q_{\text{sym}}\)), though the correlation is not very strong. Furthermore, there is no significant correlation between the delta-resonance coupling parameters (\(R_{\sigma \Delta}\), \(R_{\omega \Delta}\), \(D_{\sigma \omega}\)) and nuclear matter properties at saturation density, except in a few cases where \(R_{\sigma \Delta}\) exhibits a strong negative correlation with the slope parameter \(L\) and the kurtosis of the symmetry energy (\(Z_{\text{sym}}\)). 

Additionally, nuclear matter properties at saturation density do not exhibit a significant correlation with neutron star observables. The individual delta-resonance couplings (\(R_{\sigma \Delta}\), \(R_{\omega \Delta}\)) also do not show meaningful correlations with neutron star properties. However, the parameter \(D_{\sigma \omega}\) and the relative strength of the delta potential to the nucleon potential (\(U_{\Delta}/V_N\)) significantly influence neutron star properties in delta-resonance-admixed neutron stars. Specifically, central energy density (\(E_c\)), pressure (\(P_c\)), and density (\(\rho_c\)) are negatively correlated with \(D_{\sigma \omega}\) and \(U_{\Delta}/V_N\), whereas the maximum mass radius (\(R_{\text{max}}\)), tidal deformability at maximum mass (\(\Lambda_{\text{max}}\)), and the radius and tidal deformability for neutron stars with mass 1.4 \(M_\odot\) are strongly positively correlated. This highlights the relative difference between \(R_{\sigma \Delta}\) and \(R_{\omega \Delta}\), encapsulated in \(D_{\sigma \omega}\), as a defining parameter in delta-resonance-admixed neutron stars.

While many systematic and well-established correlations among various neutron star properties are evident, we observe several important and notable results. There is a very strong correlation between the fundamental mode frequency at 1.4 \(M_\odot\) (\(f_{1.4}\)) and the radius at 1.4 \(M_\odot\) (\(R_{1.4}\)), with a correlation coefficient of \(-0.94\). Similarly, \(f_{1.4}\) is strongly correlated with the tidal deformability at 1.4 \(M_\odot\) (\(\Lambda_{1.4}\)), exhibiting a correlation coefficient of \(-0.98\).  These trends are also visible in their marginalized posterior distributions, as shown in Fig.~\ref{fig:tov_post} and it can be seen that the relationship is linear. A similar correlation was reported in \cite{guha2023, Kumar_2023}, although we find an even stronger correlation among these pairs when delta resonances are included in neutron star matter. Additionally, \(f_{1.4}\) is strongly correlated with maximum mass neutron star properties, such as the radius corresponding to the maximum mass (\(R_{\text{max}}\)) and the tidal deformability at maximum mass (\(\Lambda_{\text{max}}\)). The pressure (\(p1_{1.4}\)) mode frequency exhibits a similar trend to the fundamental (\(f\)) mode frequency, although the correlation strength is slightly lower.  Similar to the \(f\)-mode, the damping time of the fundamental mode at 1.4 \(M_\odot\) (\(\tau_{f_{1.4}}\)) is strongly correlated with both \(R_{1.4}\) and \(\Lambda_{1.4}\), but with a positive correlation. In contrast, the damping time of the pressure mode (\(\tau_{p_{1.4}}\)) exhibits a strong negative correlation with \(R_{1.4}\) and \(\Lambda_{1.4}\).

The relationship between the \(f\)-mode oscillations and the stellar radius is crucial due to its connection between two different channels of neutron star measurements, namely gravitational waves (GW) and X-ray observations. In this context, it is valuable to establish empirical relations for the pairs \(f_{1.4} - R_{1.4}\) and \(f_{1.4} - \Lambda_{1.4}\). To quantify these relationships, we fit the data using the empirical formula \cite{guha2023, Kumar_2023}:
\begin{equation}
    f_{1.4} = a R_{1.4} + b
\end{equation}

based on the ensemble of equations of state (EoS) from our posterior distributions. The fitted coefficients for these relations are provided in Table~\ref{tab:fm_fit}.

\begin{table}[h]
    \centering
    \renewcommand{\arraystretch}{1.2} % Adjust row spacing
    \setlength{\tabcolsep}{12pt} % Adjust column spacing
    \caption{Comparison of the fitted parameters (\(a\) and \(b\)) for the empirical relation \( f = a R_{1.4} + b \) obtained in this study and previous studies. The frameworks used in each study are also indicated.}
    \begin{tabular}{lcc}
        \hline
        \hline
        \textbf{Study (Framework)} & \boldmath\(a\) & \boldmath\(b\) \\
        \hline
        Current Study (FGR) & -0.1989 & 4.2108 \\
        Guha \textit{et al.} \cite{guha2023} (FGR) & -0.1933 & 4.1949 \\
        Kumar \textit{et al.} \cite{Kumar_2023} (Cowling) & -0.22 & 5.1 \\
        \hline
    \end{tabular}
    \label{tab:fm_fit}
\end{table}
A remarkable consistency is observed in the values of \(a\) and \(b\) between this study and the work of Guha \textit{et al.}, both of which employ the FGR framework. In contrast, Kumar \textit{et al.} \cite{Kumar_2023} utilized the Cowling approximation, which is known to introduce deviations of up to 10–30\%. This strong agreement suggests that the empirical relation between the \(f\)-mode frequency and the stellar radius may be a fundamental and universal relation. Notably, this relation holds true regardless of the underlying composition of the neutron star. While the present study incorporates a delta-resonance-admixed hypernuclear EoS, previous works primarily considered nucleonic EoSs.  This universal nature of the \(f\)-mode and radius relation has profound implications for neutron star astrophysics, offering a powerful tool for constraining stellar properties from gravitational wave and X-ray observations. In the \(\Delta\)-admixed neutron star with the DDRH model, one can see from Fig. \ref{fig:corr} and Fig. \ref{fig:tov_post} that \(D_{\sigma \omega}\) significantly impacts the global properties of the neutron star. While a greater value of \(D_{\sigma \omega}\) corresponds to a lower radius and tidal deformability, it also leads to a higher $f$-mode frequency. While Ref. \cite{Pradhan_2022} estimated a strong correlation between the $f$-mode frequency and the effective nucleon mass \(M^*\), we see no such correlation in our study. Furthermore, we did not find any strong positive or negative correlation between non-radial oscillation mode frequencies and nuclear matter saturation properties. Ref.~\cite{Thapa_2023} reported some correlations among pairs such as the $f$-mode and $K_{\text{sat}}$, $Q_{\text{sat}}$, and $K_{\text{sym}}$. Ref.~\cite{Thapa_2023} reported some correlations among pairs such as the $f$-mode and $K_{\text{sat}}$, $Q_{\text{sat}}$, and $K_{\text{sym}}$. However, as evident from their Fig.~3, the correlations are not particularly strong.

%\textbf{do we also need p1 plot like for f mode?? Vivek?}

Finally, in Fig. \ref{fig:cs}, we present the speed of sound squared (\(C_s^2\)), trace anomaly (\(\delta\)), and conformality factor (\(d_c\)) for the ensemble of EoS in our posterior, incorporating the \(\Delta\) resonance in hypernuclear matter. The speed of sound is a fundamental quantity that provides crucial insights into the interior composition of neutron stars. 
The right panel of Fig. \ref{fig:cs} illustrates the 90\% CI for the speed of sound squared as a function of baryon density. A prominent peak in the speed of sound is observed within a neutron star of mass \( M \sim 1.4 M_{\odot} \), while also displaying a broad range of values. 
%This behavior is primarily attributed to the presence of \(\Delta\) resonances as nuecleonic or hypersonic eos do not show such a large bump in the speed of sound \cite{Malik_2022_1, Malik_2022}.  \textcolor{blue}{The peak in the speed of sound corresponds to the transition region where the $\Delta$ population is increasing rapidly. At this point, the matter undergoes a change in composition that temporarily enhances the pressure response to density, thereby resulting in a local maximum in the speed of sound. This behavior is especially pronounced due to the first-order nature of the phase transition involving the $\Delta^-$, as discussed in our treatment of spinodal instabilities and the Maxwell construction (see Section III and Fig. 4). Physically, this peak signals a transient stiffening of the EOS caused by the rapid rearrangement of the matter composition as $\Delta$ baryons, especially the $\Delta^-$ and $\Delta^0$, start contributing significantly to the pressure. Once this transition settles and other hyperons appear, the speed of sound begins to decrease again due to further softening of the EOS.}  
This behavior is primarily attributed to the presence of \(\Delta\) resonances
(as shown in the particle fraction plot (Fig.~\ref{fig:pf}))
as nucleonic or hyperonic EOS do not show such a large bump in the speed of sound \cite{PhysRevC.105.015802, Malik_2022_1, Malik_2022}. 
The emergence of these states leads to a rapid change in the composition of dense matter, inducing a non-monotonic behavior in the pressure response. This results in a temporary stiffening of the EOS, reflected as a local maximum in the speed of sound. Such a peak arises naturally in our framework due to the first-order nature of the phase transition involving $\Delta$-baryons and the associated spinodal instability treated via Maxwell construction. 
Importantly, this behavior may serve as a distinctive signature of $\Delta$ resonance dynamics in dense matter. 
In the context of neutron star mergers or gravitational wave asteroseismology, where the speed of sound plays a central role in determining post-merger oscillation frequencies and spectral features, the presence of such a peak could imprint observable consequences. Therefore, identifying and interpreting such features in gravitational wave data from future high-precision observations could offer an indirect probe of the onset and role of $\Delta$-resonances in neutron star interiors.
Notably, the peak in the speed of sound aligns well with results reported in the literature \cite{Marczenko_2023, Takatsy_2023} using the  ensemble of equations of state that fulfill multimessenger constraints.

Recently, the trace anomaly, defined as \(\delta = \frac{1}{3} - \frac{P}{E}\), has been proposed as a measure of conformality in neutron stars \cite{Annala2023} and efforts are underway to better understand its behavior and implications for dense matter physics \cite{Chatterjee:2023ecc}. As matter approaches the conformal limit, the value of \(\delta\) tends to zero. In our case, within the \(\Delta\)-resonance admixed hyperonic neutron star scenario, \(\delta\) consistently approaches zero from above. Literature suggests that in some cases, the conformal limit can be reached both from above and below \cite{Marczenko_2023}. Another recent study introduced an alternative measure of conformality by combining the trace anomaly with its logarithmic derivative \cite{PhysRevLett.129.252702}. The conformality factor is defined as 
$d_c = \sqrt{\delta^2 + (\delta')^2}$,  where \(\delta' = \frac{d\delta}{d \ln E}\). This measure has been proposed based on the observation that hadronic equations of state (EoS) can be distinguished from conformal systems using \( d_c \).  It has been further conjectured that conformal matter can be identified by the criterion \( d_c < 0.2 \), which is represented by the blue dashed line in Fig. \ref{fig:cs} \cite{Annala2023}. 
Although our posterior for \( d_c \) lies well within the range proposed by Annala \textit{et al.} \cite{Annala2023}, it crosses the \( d_c = 0.2 \) threshold just beyond the canonical neutron star mass, even without  incorporating a quark matter phase transition. Examining our posterior distributions, we observe that they exhibit hybrid EoS kind of  behavior and do not comply with this conjecture reported in \cite{Annala2023}. These results suggest that the properties proposed  for identifying deconfined matter are not universally unique but rather model-dependent, warranting further investigation. A similar behavior has been observed in certain purely nucleonic EoS, such as the FSU2R EoS \cite{Takatsy_2023}.  Additionally, a noticeable bump appears at low densities in \( d_c \), a feature that is typically absent in purely hyperonic or nucleonic EoS. This anomaly arises from the fact that many EoS exhibit a first-order phase transition associated with the onset of \(\Delta\) resonances at low densities—a behavior reminiscent of the hadron-quark phase transition. We use Set I for this analysis and have verified that Set II follows the same trend.

\section{\label{summary} Summary}
In this work, we present a Bayesian analysis of neutron star matter incorporating $\Delta$-resonances within a density-dependent relativistic hadron (DDRH) framework. Our study systematically investigates the role of $\Delta$-baryons in shaping the equation of state, neutron star observables, and the fundamental properties of dense matter, leveraging constraints from both nuclear physics and multi-messenger astrophysical observations. Using Bayesian inference, we constrain the model parameters by integrating nuclear matter saturation properties, theoretical constraints from chiral effective field theory ($\chi$EFT), and astrophysical data, including pulsar mass–radius measurements from NICER and tidal deformability limits from the GW170817 event. To assess the sensitivity of our findings to nuclear matter properties, we consider two sets of constraints that differ in the assumed range of symmetry energy, providing a comprehensive analysis of $\Delta$-baryon effects in neutron star matter.

Our findings reveal that the inclusion of $\Delta$-resonances leads to a significant softening of the EoS at low densities while maintaining sufficient stiffness at high densities to support neutron stars exceeding $2M_{\odot}$.  Furthermore, the presence of $\Delta$-resonances effectively satisfies the constraints imposed by GW170817 on the EoS and tidal deformability. Notably, $\Delta$-resonances contribute to a more compact neutron star configuration, leading to a reduced radius.   Our analysis does not favor high values of the symmetry energy $J$ and its slope $L$ when incorporating the PREX-II constraints, a tension that has also been reported in the literature. Among the $\Delta$-resonances, $\Delta^-$ and $\Delta^0$ are found to appear before the $\Lambda$ hyperon, suggesting that they can exist in the outer layers of the neutron star core. Since these particles populate the outer core, they might influence the dynamics of neutron star mergers, potentially affecting the tidal deformability, mass ejection, and post-merger oscillations, which could have observable consequences in gravitational wave signals.

Our Bayesian analysis provides robust statistical constraints on the coupling strengths of $\Delta$-resonances to mesonic fields, revealing that in the DDRH framework, the relative strength of the scalar and vector couplings, characterized by the parameter $D_{\sigma \omega}$, plays a crucial role in determining neutron star properties. We find that, while we fix $R_{\rho \Delta}$=1, values of $R_{\sigma \Delta}$ and $R_{\omega \Delta}$ greater than 1 are preferred, with $R_{\omega \Delta}$ lying in the range $[1.07 - 1.3]$,  and the difference $R_{\sigma \Delta} - R_{\omega \Delta}$ favored in the range $[0.06 - 0.29]$, when adhering to the currently accepted symmetry energy range of $30 - 33$ MeV. Correlation analyses further indicate that $D_{\sigma \omega}$ significantly influences neutron star radii, tidal deformability, and $f$-mode oscillation frequencies. Furthermore, we study the impact of $\Delta$-resonances on neutron star cooling via the direct Urca (DU) process.  While it is well known that $\Delta$-resonances lower the DU threshold for neutrons, we find that the $\Delta^-$ and $\Lambda$ DU processes have approximately the same threshold density and are favored to occur at the central density of a canonical neutron star. This could have a significant impact on neutron star cooling.

We systematically investigate the influence of $\Delta$-resonances on neutron star observables, including the mass-radius ($M$–$R$) relation, tidal deformability, and non-radial oscillation modes. The fundamental ($f$-mode) and first pressure ($p_1$-mode) oscillation frequencies are computed within full general relativity, providing new empirical relations that link $f$-mode frequencies to stellar compactness. We find a strong correlation between the $f$-mode frequency and neutron star radius, reinforcing the potential of gravitational wave asteroseismology in probing the internal structure of neutron stars.  Additionally, we analyse empirical linear relationship between the $f$-mode frequency at $1.4M_{\odot}$ ($f_{1.4}$) and the corresponding radius ($R_{1.4}$), demonstrating that this relation remains valid with a similar fitting coefficient even when comparing neutron stars with different internal compositions. A more compact neutron star, resulting from the presence of $\Delta$-resonances, leads to a slightly higher $f$-mode frequency. Additionally, we analyze the speed of sound squared ($C_s^2$), the trace anomaly ($\delta$), and the conformality factor ($d_c$) to explore deviations from conformal behavior in neutron star matter. Our results indicate that the presence of $\Delta$-baryons leads to a characteristic peak in the speed of sound and a model-dependent behavior of the conformality factor, which crosses the $d_c = 0.2$ threshold without invoking a quark phase transition. This suggests that the criteria proposed in previous literature for identifying deconfined quark matter are not universally unique and remain  model-dependent.

In conclusion, our study highlights the astrophysical relevance of $\Delta$-resonances in neutron stars and their potential role in shaping the dense matter EoS. By integrating nuclear theory with multi-messenger astrophysical constraints, we provide new constraints on the dense matter EoS while considering $\Delta$-resonances as additional degrees of freedom. 

\section*{Acknowledgement}
MS acknowledges the financial support from the Science and
Engineering Research Board, Department of Science andTechnology, Government of India through Project
No. CRG/2022/000069 and the HPC cluster facility at IIT Jodhpur. 
VBT acknowledges the partial support provided by the IUCAA Associateship Programme to carry out a part of this work at IUCAA.

\bibliographystyle{apsrev4-2}

\bibliography{main}% Produces the bibliography via BibTeX.

\end{document}